\PassOptionsToPackage{a4paper,margin=1in}{geometry}
\documentclass{osa-article}

\usepackage{amsmath} 

\newtheorem{theorem}{Theorem}[section]
\newtheorem{proposition}[theorem]{Proposition}
\newtheorem{corollary}[theorem]{Corollary}
\newtheorem{definition}[theorem]{Definition}
\newtheorem{lemma}[theorem]{Lemma}
\newtheorem{remark}[theorem]{Remark}

\newenvironment{proof}{\par\noindent\textit{Proof.}\ }{\hfill$\square$\par}

\usepackage{graphicx}
\usepackage{hyperref}

\usepackage{titlesec}
\usepackage{float}
\usepackage{dsfont}

\usepackage{graphicx}
\usepackage{hyperref}
\usepackage{enumitem}
\usepackage{cite}
\usepackage{tikz}
\usepackage{bm}
\usepackage{booktabs}
\usepackage{array}
\usepackage{framed}
\usepackage{xcolor}
\usepackage{physics}

\newcommand{\figref}[1]{Fig.~\ref{#1}}
\titleformat{\paragraph}
{\normalfont\normalsize\bfseries}{\theparagraph}{1em}{}
\titlespacing*{\paragraph}
{0pt}{3.25ex plus 1ex minus .2ex}{1.5ex plus .2ex}

\journal{osac}

\articletype{Research Article}

\begin{document}
\title{The Role of Fractional Dimension in  Study Physics: A Two-Channel Representation with Geometric Memory}
{\centering{Ali Dorostkar\\Fractal Group, Isfahan, Iran\\m110alidorostkar@gmail.com\\}}



\begin{abstract}
\noindent This study has three parts. In part A , we explore the field of physics through the lens of fractional dimensionality. We propose that space is not confined to integer dimensions alone, but can also be understood as a superposition of spaces that exist between these integer dimensions. The concept of fractional dimensional space arises from the idea that the space between integer dimensions is filled, which occurs through the application of a fractional derivative operator (the local part) that rotates the integer dimension to encompass all spaces between two integers.\\
It examines how fractional dimensional frameworks can enhance our understanding of classical mechanics, particularly regarding the duality of memory versus no-memory behavior, or local versus non-local dynamics. In the lens of  fractional dimension, motion in classical physics can be analyzed through two distinct solutions. When the fractional dimensional trajectory \( \alpha \) takes values of 1 or 2—corresponding to the first and second integer derivatives— it represents linear and accelerated systems, respectively, yielding trivial solutions via differentiation. However, if the fractional dimensional trajectory \( \alpha \) evolves as a linear  function of time in fractional dimensional space   (e.g. \( \alpha \approx -2.63 t + 2.83\)) or follows a nonlinear path, surprisingly it results in a non-trivial solution for linear and accelerated systems, respectively.
This approach offers a broader framework for describing motion, extending to memory (non-local) effect beyond traditional local integer-order differentiation .  Moreover, we propose that the coupling of space and time, commonly referred to as space-time, is better understood as space-dimension-time within this framework, where the dimension serves as an interconnecting platform.\\

\noindent In part B, the future works and extension of memory geometry to broader physics area such as  viscoelasticity, Electromagnetism, quantum mechanics have been arranged. For instance, we derive the wave equation from the perspective of fractional dimensions, focusing on linear trajectories in fractional dimensional space. This approach provides new insights into the behavior of electromagnetic waves in both lossless and lossy media, and it offers a fresh interpretation of the Doppler effect and gravitational redshift (or blueshift) phenomena from the standpoint of fractional dimensionality.\\
\\
In part C, we build the mathematical framework required for part A (physics study) in fractional dimensional space. 
\end{abstract}

\textbf{\Large {Part A: Physics Study}}\\
\section{Introduction}
Understanding the nature of reality is a fundamental pursuit in science, particularly in the study of physics. However, according to Gödel's incompleteness theorems \cite{smullyan1992godel}, achieving a complete and absolute understanding of nature is inherently impossible. These theorems suggest that no, single comprehensive theory can fully encapsulate the entirety of natural phenomena. As a result, scientific progress is characterized by the continuous refinement of models, with each iteration providing a closer approximation to reality, though never reaching perfection. This iterative process has been a defining feature in the history of physics.

Classical, or Newtonian, physics has long served as a foundational framework, effectively describing a wide range of natural phenomena. It has proven to be a powerful tool for explaining the dynamics of macroscopic objects and most observable occurrences in the natural world. However, classical physics faces limitations when addressing the dynamics of particles at microscopic scales or providing a fundamental explanation for gravitational forces. These challenges have driven the development of more advanced theories, leading to the advent of general relativity (GR) and quantum mechanics (QM).

Einstein's general relativity, for instance, redefines gravity not as a force but as a consequence of the curvature of space-time induced by mass, thus offering a framework for understanding large-scale physical phenomena \cite{wald2010general}. On the other hand, quantum mechanics provides a robust model for describing the behavior of particles at very small scales \cite{galindo2012quantum}. Despite their successes, both theories still grapple with unresolved issues—quantum mechanics, for example, struggles with the concept of particle trajectories and the non-locality effect. Meanwhile, general relativity does not easily reconcile with quantum principles.

Numerous efforts have been made to unify these two pillars of modern physics, resulting in theories such as M-theory and string theory \cite{becker2006string, holstein2006graviton}. Despite these efforts, a complete connection between quantum mechanics and gravity remains elusive, highlighting the persistent gaps in our understanding of the natural world.

This ongoing quest for deeper understanding reminds us that there will always be open questions in physics, driving the continuous evolution of our knowledge of nature. In this context, our study seeks to explore how additional insights can be gained by considering physics from a fractional dimensional perspective. To do so, we first establish several foundational axioms and then develop a conceptual framework that examines how the movement of particles in integer-dimensional space corresponds to movement in fractional-dimensional space.\\
\noindent For example, in traditional physics, the three axes $x_1, x_2, x_3$ define a space of three integer dimensions. However, from our perspective, it is essential to also account for fractional dimensions. As demonstrated in \cite{dorostkar2022fourier}, fractional basis vectors can be conceptualized as rotations of the basis vectors of integer dimensions. This rotation is mathematically represented by the fractional derivative (FDr) operator, which effectively fills the space between integer dimensions, thereby creating a fractional-dimensional space. By establishing several axioms, we construct a mathematical framework using the local FDr operator, allowing us to investigate the moving (propagation) of particles (or signals) along specific trajectories within this fractional-dimensional space.

Memory and nonlocality play a central role in a wide range of physical systems,
from viscoelastic materials and anomalous diffusion to complex media and
effective field theories \cite{mainardi2010fractional,metzler2000random,bagley1983theoretical}.
Fractional calculus has emerged as a powerful mathematical tool for modeling
such effects, as fractional derivatives and integrals naturally encode history
dependence through nonlocal kernels \cite{podlubny1999fractional,kilbas2006theory}.
Despite its success, most applications of fractional calculus in physics adopt
a fixed fractional order, typically introduced phenomenologically and tuned to
experimental data. As a result, the fractional order often lacks a direct
geometric or dynamical interpretation \cite{tarasov2011fractional}.

In parallel, classical mechanics is traditionally formulated in terms of
local-in-time differential equations, where the state of a system is defined at
an instant and evolution is governed by integer-order derivatives
\cite{goldstein2002classical}. This locality obscures the possibility that
classical trajectories may admit nontrivial internal structure associated with
memory, even when the observable motion remains strictly Newtonian. While
memory effects are sometimes introduced through additional degrees of freedom
or constitutive relations, they are not usually interpreted as intrinsic
geometric features of kinematics.

The present work aims to bridge this gap by introducing a
fractional--dimensional kinematic framework in which the fractional order is
treated as a dynamical descriptor of motion rather than a fixed parameter.
Specifically, we promote the fractional order to a time-dependent order
trajectory $\alpha(t)$ and interpret it as a geometric measure of memory
embedded within the evolution of the system. This approach does not modify the
observed trajectory $x(t)$; instead, it reveals an internal representation in
which the same trajectory can be realized through either a purely local
(integer-order) description or a nonlocal (fractional-order) description.

To formalize this idea, we introduce a two-channel decomposition of motion into
a local channel governed by integer-order calculus and a nonlocal channel
governed by fractional operators. The relative contribution of these channels
is controlled by convex weights, while the order trajectory $\alpha(t)$
determines the structure of the memory channel. Within this framework,
Newtonian mechanics emerges as a special, trivial branch corresponding to
integer-order limits.

A key result of this work is that imposing kinematic consistency conditions
selects admissible order trajectories without introducing an independent
equation of motion for $\alpha(t)$. Uniform motion is shown to require a
constant memory slope, whereas uniform acceleration requires a linearly
evolving memory slope. This establishes a direct geometric interpretation of
acceleration as curvature in the memory channel, even when the physical
trajectory remains quadratic in time. The framework is further illustrated
through the harmonic oscillator, where sinusoidal motion is sustained by a
time-dependent fractional order without altering the restoring force.

This paper is organized as follows. In Section~2, we introduce the
fractional--dimensional representation of trajectories and define the relevant
fractional operators. Section~3 formulates the  two-channel kinematics
and recovers Newtonian limits. Sections~3 and~4 analyze uniform motion,
uniform acceleration, and their associated memory slopes. Section~5 applies the
framework to the harmonic oscillator. We conclude with a discussion of the
physical interpretation, limitations, and future directions, including the
derivation of $\alpha(t)$ from first-principles dynamics. The framework should therefore be interpreted as a geometric representation of classical motion rather than a modification of Newtonian dynamics.

\section{Fractional-Dimensional Representation of Trajectories}

According to the mathematical results presented in Section~13 (Part~C),
specifically Theorems~\ref{thm:sc1} and~\ref{thm:sc2}, there exist two distinct
approaches for representing an arbitrary function by means of fractional
operators. Geometrically, the \emph{fractional--dimensional tangent line}
(Fig.~\ref{TL1}) connects the point $(t_0,f(t_0))$ with fractional dimension
$\alpha_0$ to another point $(t_n,f(t_n))$ with dimension $\alpha_n$, such that
\begin{equation}
(\mathcal{O}^{\alpha_0}f)(t_0)
=
(\mathcal{O}^{\alpha_n}f)(t_n),
\end{equation}
where $\mathcal{O}^{\alpha}$ denotes a fractional operator (derivative or
integral).

\medskip

We represent a function $f$ using a tangent--like affine form whose slope is
generated by a variable--order fractional operator:
\begin{equation}
f(t)
=
f(t_0)
+
(\mathcal{O}^{\alpha(t)}f)(\tau)\,(t-t_0),
\qquad
\tau\in\{t_0,t\}.
\label{eq:frac_tangent}
\end{equation}
The operator $\mathcal{O}^{\alpha}$ may be chosen as the Caputo derivative
$({}^{\mathrm{C}}D^\alpha)$, the Riemann--Liouville derivative
$({}^{\mathrm{RL}}D^\alpha)$, or the Riemann--Liouville integral
$({}^{\mathrm{RL}}I^\alpha)$.

To avoid ambiguity when the order crosses zero, we adopt the convention
\begin{equation}
\mathcal{O}^{\alpha}
:=
\begin{cases}
{}^{\mathrm{C}}D^{\alpha}, & \alpha>0,\\[4pt]
\mathrm{Id}, & \alpha=0,\\[4pt]
{}^{\mathrm{RL}}I^{-\alpha}, & \alpha<0,
\end{cases}
\label{eq:O_def}
\end{equation}
so that positive order corresponds to differentiation and negative order to integration. This convention ensures analytic continuity of the operator across positive and negative orders.

The variable order $\alpha(t)$ encodes how the system interpolates between local
(instantaneous) behavior and nonlocal (memory-dependent) behavior, acting as a
fingerprint of the underlying geometry and dynamics.

Recalling the classical Taylor expansion about $t_0$, we introduce a convex
mixture of local and nonlocal representations:
\begin{equation}
f(t)
=
C_1
\left(
\sum_{n=0}^{\infty}
\frac{f^{(n)}(t_0)}{n!}(t-t_0)^n
\right)
+
C_2
\left(
f(t_0)
+
(\mathcal{O}^{\alpha(t)}f)(\tau)\,(t-t_0)
\right),
\label{eq:two_channel_f}
\end{equation}
where $C_1$ and $C_2$ satisfy
\begin{equation}
C_1+C_2=1,
\qquad
C_1,C_2\ge 0.
\end{equation}
The first term represents the local  (memoryless) contribution, while the second
term represents the nonlocal (memory) contribution.
So we may write succinctly
\begin{equation}
f(t)=C_1 f_{\mathrm{loc}}(t)+C_2 f_{\mathrm{mem}}(t).
\label{eq:trajectory_two_channel}
\end{equation}

\subsection{Fractional Calculus for Power Functions}

For power functions $(t-t_0)^n$, the Caputo derivative satisfies~\cite{RG}
\begin{equation}
{}^{\mathrm{C}} D^{\alpha}_{t_0} (t - t_0)^n =
\begin{cases}
\dfrac{\Gamma(n + 1)}{\Gamma(n - \alpha + 1)} \,(t - t_0)^{\,n - \alpha},
& 0<\alpha \leq n, \\[8pt]
0, & \alpha > n,
\end{cases}
\label{eq:caputo-power}
\end{equation}
which is especially convenient in physical modeling because the fractional
derivative of a constant is zero.

For the Riemann--Liouville integral, one has
\begin{equation}
{}^{\mathrm{RL}} I^{\alpha}_{t_0} (t - t_0)^n
=
\frac{\Gamma(n + 1)}{\Gamma(n + \alpha + 1)} (t - t_0)^{n +\alpha},
\qquad \alpha>0.
\label{eq:RLint-power}
\end{equation}
Using the convention \eqref{eq:O_def}, these combine into the compact
expression
\begin{equation}
\boxed{
\mathcal{O}^{\alpha}_{t_0} (t - t_0)^n
=
\frac{\Gamma(n + 1)}{\Gamma(n - \alpha + 1)} (t - t_0)^{n -\alpha},
\qquad \alpha\in\mathbb{R},\ \alpha\le n,
}
\label{eq:O_power_compact}
\end{equation}
where $\alpha>0$ corresponds to the Caputo derivative and $\alpha<0$ corresponds
to the Riemann--Liouville integral.
The identity \eqref{eq:O_power_compact} is understood away from poles of
$\Gamma(n-\alpha+1)$ (i.e.\ where $n-\alpha+1\in\{0,-1,-2,\dots\}$). Moreover, for
$\alpha>n$ the Caputo derivative of $(t-t_0)^n$ vanishes, consistent with
\eqref{eq:caputo-power}. In the negative-order sector ($\alpha<0$), the same
Gamma ratio coincides with the Riemann--Liouville integral rule
\eqref{eq:RLint-power} under the convention \eqref{eq:O_def}.
\subsection{Dimensionally normalized operators}

To maintain dimensional consistency under variable order $\alpha(t)$, introduce
a reference time scale $T_0>0$ (e.g.\ $T_0=1~\mathrm{s}$). We define normalized
operators for velocity and acceleration units:
\begin{equation}
\mathcal{O}^{\alpha(t)}_{\mathrm{sc}}
:=T_0^{\alpha(t)-1}\,\mathcal{O}^{\alpha(t)},
\label{eq:scaled_ops}
\end{equation}
Let $\hat t=(t-t_0)/T_0$. Then, for powers,
\begin{equation}
\mathcal{O}^{\alpha(t)}_{\mathrm{sc}}(t-t_0)^n
=
T_0^{n-1}\,
\frac{\Gamma(n+1)}{\Gamma(n+1-\alpha(t))}\,
\hat t^{\,n-\alpha(t)}.
\label{eq:scaled_power_v}
\end{equation}
For a position variable $x$ with units of length, the normalization in
\eqref{eq:scaled_ops} is chosen so that $\mathcal{O}_{\mathrm{sc}}^{\alpha(t)}x$
has the physical dimension of velocity for any order $\alpha(t)$.
This is why it naturally appears as a ``memory slope'' in the
two-channel trajectory representation. All kinematic consistency
conditions are therefore imposed on the scaled operator
$\mathcal{O}_{\mathrm{sc}}^{\alpha(t)}$. 

\section{ Two-channel kinematics and Newtonian limits}
Equation~\eqref{eq:trajectory_two_channel} shows that motion may be understood as the
superposition of two kinematic channels: a local channel governed by integer-order
calculus and a nonlocal channel governed by fractional operators.
The order trajectory (at the relevant kinematic level) acts as a kinematic order
trajectory that determines the relative influence of memory on the motion.

In general, the admissible fractional order trajectory depends on the kinematic
level at which constraints are imposed. We therefore distinguish between order
functions associated with velocity constraints, denoted by $\alpha_v(t)$, and
those associated with acceleration constraints, denoted by $\alpha_a(t)$.
These order trajectories are selected implicitly by kinematic consistency
conditions rather than by independent evolution equations.\\

From a physical perspective, the two-channel decomposition suggests
that a single observable trajectory may admit both a local (memoryless)
description and a nonlocal (memory-dependent) internal representation.
We interpret the order trajectory $\alpha(t)$ as encoding geometric memory.

\subsection{Two-channel representation of uniform motion}
A body remains at rest, or moves with constant velocity \(v\) along a straight trajectory,
unless acted upon by a net external force. In one spatial dimension, uniform motion is described
by the classical Newtonian trajectory
\begin{equation}
x(t)=v\,(t-t_0)+x(t_0).
\label{eq:xN}
\end{equation}

Starting from the two-channel representation \eqref{eq:two_channel_f} and applying it to the position
function \(x(t)\), we rewrite the trajectory as a convex superposition of a local (memoryless)
channel and a nonlocal (memory) channel:
\begin{equation}
x(t)=
C_1\Big(v(t-t_0)+x(t_0)\Big)
+
C_2\Big((\mathcal{O}_{\mathrm{sc}}^{\alpha_v(t)}x)(\tau)\,(t-t_0)+x(t_0)\Big),
\qquad C_1+C_2=1,
\label{eq:traj_axiom1}
\end{equation}
where \(\tau\in\{t_0,t\}\) specifies the anchoring strategy, and
\(\mathcal{O}^{\alpha_{v}(t)}_{\mathrm{sc}}\) is a unit-normalized fractional operator (so that it has
the physical dimension of inverse time). By collecting terms in \eqref{eq:traj_axiom1}, we obtain
\begin{equation}
x(t)=
\Big(C_1\,v + C_2\,(\mathcal{O}_{\mathrm{sc}}^{\alpha_v(t)}x)(\tau)\Big)\,(t-t_0)+x(t_0).
\label{eq:traj_collected}
\end{equation}
Comparing \eqref{eq:traj_collected} with the Newtonian form \eqref{eq:xN} shows that the
constant velocity \(v\) can be interpreted as a \emph{two-channel slope}, 
Motivated by the coexistence of a trivial order branch $\alpha_{v}(t)=1$ (local, memoryless)
and a nontrivial order trajectory $\alpha_{v}(t)\neq 1$ (nonlocal, memory). We introduce the local and memory velocity components
\begin{equation}
v_{\mathrm{loc}}(t):=\frac{dx(t)}{dt},
\qquad
v_{\mathrm{mem}}(t):=(\mathcal{O}_{\mathrm{sc}}^{\alpha_v(t)}x)(\tau),
\end{equation}
and model the physical velocity as their convex combination,
\begin{equation}
\boxed{
v
=
C_1\,v_{\mathrm{loc}}(t)
+
C_2\,v_{\mathrm{mem}}(t).
}
\label{eq:two_channel_velocity_axiom1}
\end{equation}
To obtain a nontrivial memory branch (with \(C_2\neq 0\)), we impose the natural consistency
condition that the memory channel reproduces the same physical velocity:
\begin{equation}
\boxed{
(\mathcal{O}_{\mathrm{sc}}^{\alpha_v(t)}x)(\tau)=v.
}
\label{eq:velocity_constraint_axiom1}
\end{equation}
Equation \eqref{eq:velocity_constraint_axiom1} is the \emph{consistency condition}
for the velocity-level order trajectory $\alpha_v(t)$: it constrains the admissible
order trajectories $\alpha_v(t)$ (for a chosen anchor $\tau$) such that the
fractional-dimensional slope matches the observed constant velocity.
The Newtonian (memoryless) description is recovered either by switching off the
memory channel ($C_2=0$), or by the degenerate case in which the memory channel
reduces to the local one (e.g.\ $\alpha_v(t)\equiv 1$ so that
$\mathcal{O}_{\mathrm{sc}}^{\alpha_v}$ coincides with the first derivative, up to the chosen
normalization). These two limits are conceptually distinct: $C_2=0$ removes the
memory channel, whereas $\alpha_v\equiv 1$ keeps it but collapses it to an
integer-order operator.
The nontrivial branch corresponds to $C_2\neq 0$, where the same uniform motion is
reproduced through a nonlocal operator with variable order $\alpha_v(t)$, thus encoding
memory effects without changing the observable trajectory $x(t)$.\\

\begin{proposition}[Uniform motion constraint on the memory slope]
\label{prop:uniform_memory_slope}
Consider one--dimensional uniform motion with constant physical velocity $v$,
described by the Newtonian trajectory
\begin{equation}
x(t)=v\,(t-t_0)+x(t_0).
\end{equation}
Assume a two--channel fractional--dimensional representation of the form
\begin{equation}
x(t)=x(t_0)+C_1 v\,\Delta t+C_2 A(t)\,\Delta t,
\qquad
\Delta t:=t-t_0,
\end{equation}
where $C_1+C_2=1$, $C_1,C_2\ge 0$, and
\(
A(t):=(\mathcal{O}_{\mathrm{sc}}^{\alpha_v(t)}x)(\tau)
\)
denotes the memory slope.
\end{proposition}

\begin{proof}
Differentiation yields the velocity
\begin{equation}
v(t)=\frac{dx}{dt}
=
C_1 v
+
C_2\big(A(t)+\Delta t\,A'(t)\big),
\end{equation}
and the acceleration
\begin{equation}
a(t)=\frac{d^2x}{dt^2}
=
C_2\big(2A'(t)+\Delta t\,A''(t)\big).
\end{equation}
Uniform motion requires $a(t)=0$. For $C_2\neq 0$ this implies the differential
equation
\begin{equation}
2A'(t)+\Delta t\,A''(t)=0.
\end{equation}
The general solution is
\(
A(t)=A_0-\dfrac{K}{\Delta t}.
\)
Regularity at $t=t_0$ enforces $K=0$, hence $A(t)=A_0$ is constant. Substitution
into the velocity expression yields
\(
v=C_1 v+C_2 A_0,
\)
and therefore $A_0=v$.
\end{proof}

\begin{corollary}[Uniform motion]
Uniform motion is admissible if and only if the memory slope is constant and
equal to the physical velocity,
\begin{equation}
A(t)\equiv v.
\end{equation}
\end{corollary}
To reproduce a prescribed constant velocity $v$ via the memory channel, one may \emph{select}
$\alpha_{v}(t)$ such that $(\mathcal{O}_{\mathrm{sc}}^{\alpha_v(t)}x)(\tau)=v$.\\

\subsection{Kinematic energy in fractional-dimensional representation}
Accordingly, we define the fractional--dimensional kinetic energy as
\begin{equation}
\boxed{
E_{\alpha_{v}}(t)
\;:=\;
\frac{m}{2}
\left(
C_1v
+
C_2\,\mathcal{O}_{\mathrm{sc}}^{\alpha_{v}(t)}x(t)
\right)^2 .
}
\label{eq:fractional_energy}
\end{equation}
At this stage, $E_{\alpha_{v}}$ is introduced as a kinematic energy functional; its conservation properties will be addressed elsewhere.
Equation~\eqref{eq:fractional_energy} reduces to the classical kinetic
energy in the purely local limit $C_1=1$ and $C_2=0$. For $C_2\neq 0$,
the energy acquires an explicit dependence on the history of the motion
through the fractional operator $\mathcal{O}_{\mathrm{sc}}^{\alpha_{v}(t)}$, and
thus represents a genuinely nonlocal, memory--dependent energy.

To make the physical content of \eqref{eq:fractional_energy} explicit, we
expand the square and separate the local, nonlocal, and coupling
contributions:
\begin{align}
E_{\alpha}(t)
&=
\frac{m}{2}
\left(
C_1\,v
+
C_2\,\mathcal{O}_{\mathrm{sc}}^{\alpha_{v}(t)}x(t)
\right)^2
\nonumber\\[4pt]
&=
\underbrace{\frac{m}{2}\,C_1^2v^2}_{\displaystyle E_{\mathrm{loc}}(t)}
+
\underbrace{\frac{m}{2}\,C_2^2\left(\mathcal{O}_{\mathrm{sc}}^{\alpha_{v}(t)}x(t)\right)^2}_{\displaystyle E_{\mathrm{mem}}(t)}
+
\underbrace{m\,C_1C_2v\left(\mathcal{O}_{\mathrm{sc}}^{\alpha_{v}(t)}x(t)\right)}_{\displaystyle E_{\mathrm{int}}(t)}.
\label{eq:energy_split}
\end{align}

Here,
\begin{equation}
\boxed{
E_{\mathrm{loc}}(t)
=
\frac{m}{2}\,C_1^2v^2
}
\end{equation}
is the \emph{local (memoryless) kinetic energy} generated by the order--one
channel, while
\begin{equation}
\boxed{
E_{\mathrm{mem}}(t)
=
\frac{m}{2}\,C_2^2\left(\mathcal{O}_{\mathrm{sc}}^{\alpha_{v}(t)}x(t)\right)^2
}
\end{equation}
is the \emph{nonlocal (memory) kinetic energy} generated by the
variable--order fractional channel.

The third term,
\begin{equation}
\boxed{
E_{\mathrm{int}}(t)
=
m\,C_1C_2v\left(\mathcal{O}_{\mathrm{sc}}^{\alpha_{v}(t)}x(t)\right),
}
\end{equation}
is an \emph{interaction (mixing) energy} that couples the local and
nonlocal velocities. This term is absent when either channel is switched
off ($C_2=0$ or $C_1=0$), and it quantifies how instantaneous motion
interferes with history--dependent motion.\\

\noindent\textbf{Remark (consistency with the velocity constraint).}
Under the velocity-level consistency condition \eqref{eq:velocity_constraint_axiom1},
one has $\mathcal{O}_{\mathrm{sc}}^{\alpha_v(t)}x(\tau)=v$, hence
\[
C_1v+C_2\,\mathcal{O}_{\mathrm{sc}}^{\alpha_v(t)}x(\tau)=C_1v+C_2v=v,
\]
so that $E_{\alpha_v}(t)\equiv \frac12 m v^2$ even on a nontrivial order
trajectory. Nontrivial energy exchange between channels can arise only when the
memory channel is not constrained to reproduce the physical velocity exactly.

\subsection{Computation and fitting of the order trajectory $\alpha_v(t)$ for linear motion}
\label{sec:alpha_linear_fit}

In this section we describe how the variable order (the \emph{order trajectory})
$\alpha_v(t)$ is computed and then approximated by simple fitted functions for the
linear Newtonian trajectory
\begin{equation}
x(t)=v\,t+x_0,
\qquad t\in[t_1,t_2],
\label{eq:linear_traj}
\end{equation}
with fixed $t_0=0$, unit scale $T_0=1$, and the \emph{moving-anchor} choice $\tau=t$.

The velocity-level consistency condition (Axiom 1) requires that the
fractional-dimensional slope reproduces the constant physical velocity ($
\big(\mathcal{O}^{\alpha_v(t)}x\big)(t)=v$).
Using the power-function rule (Caputo for $\alpha>0$, Riemann--Liouville integral for
$\alpha<0$ via our unified convention) and linearity, one obtains
\begin{equation}
\boxed{
\frac{v}{\Gamma\!\big(2-\alpha_v(t)\big)}\,t^{\,1-\alpha_v(t)}
+
\frac{x_0}{\Gamma\!\big(1-\alpha_v(t)\big)}\,t^{-\alpha_v(t)}
=
v.
}
\label{eq:alpha_linear_implicit}
\end{equation}
Dividing by $v\neq 0$ shows that the dependence on $(v,x_0)$ enters through the ratio
$r:=x_0/v$:
\begin{equation}
\boxed{
\frac{t^{\,1-\alpha}}{\Gamma(2-\alpha)}
+
r\,\frac{t^{-\alpha}}{\Gamma(1-\alpha)}
=
1,
\qquad r=\frac{x_0}{v}.
}
\label{eq:alpha_ratio_form}
\end{equation}
Equation \eqref{eq:alpha_ratio_form} implies the symmetry
\begin{equation}
\alpha_v(t;v,x_0)=\alpha_v(t;-v,-x_0),
\label{eq:symmetry_vx0}
\end{equation}
since $r=x_0/v$ is invariant under $(v,x_0)\mapsto(-v,-x_0)$.
\subsubsection{Asymptotic derivation of the universal slope ($r=0$ case)}

A particularly important regime occurs when
\[
r=\frac{x_0}{v}=0,
\]
i.e., when $x_0=0$. In this case the implicit order equation
\eqref{eq:alpha_ratio_form} reduces to
\begin{equation}
\boxed{
\frac{t^{\,1-\alpha}}{\Gamma(2-\alpha)}=1.
}
\label{eq:U_equation}
\tag{U}
\end{equation}

We now analyze the nontrivial branch of \eqref{eq:U_equation}
for moderate to large $t$.
Numerically one observes that $\alpha(t)$ becomes negative.
We therefore introduce
\begin{equation}
\alpha=-\beta,
\qquad
\beta>0.
\end{equation}
Substituting into \eqref{eq:U_equation} gives
\begin{equation}
\boxed{
\frac{t^{\,1+\beta}}{\Gamma(2+\beta)}=1.
}
\label{eq:U_beta}
\end{equation}

\subsubsection{Stirling approximation.}
For large $\beta$ we use the Stirling expansion
\begin{equation}
\Gamma(\beta+2)
\sim
\sqrt{2\pi}\,\beta^{\beta+\frac{3}{2}} e^{-\beta}.
\end{equation}
Substituting into \eqref{eq:U_beta} and taking logarithms yields
\begin{equation}
(1+\beta)\ln t
-
\Big[
\Big(\beta+\tfrac{3}{2}\Big)\ln\beta
-\beta
\Big]
\approx 0.
\end{equation}

Keeping only the dominant exponential balance (i.e., the
$\beta\ln\beta$ and $\beta$ terms) gives
\begin{equation}
\beta(\ln t-\ln\beta+1)\approx 0.
\end{equation}
For $\beta>0$, this implies
\begin{equation}
\ln t-\ln\beta+1\approx 0,
\qquad
\Rightarrow
\qquad
\boxed{
\beta(t)\sim e\,t.
}
\end{equation}

Therefore,
\begin{equation}
\boxed{
\alpha_v(t)\sim -e\,t
\qquad (x_0=0,\; t\to\infty).
}
\label{eq:alpha_asymptotic}
\end{equation}

\subsection{Numerical procedure}
On the interval $t\in[1,10]$ the computed order trajectory is extremely well
approximated by a linear function,
\begin{equation}
\boxed{
\alpha_v(t)\approx m(v,x_0)\,t+b(v,x_0),
\qquad t\in[1,10].
}
\label{eq:alpha_linear_fit}
\end{equation}
We determine $m$ and $b$ by least squares regression over the computed samples
$\{(t_k,\alpha_v(t_k))\}$ and report the coefficient of determination $R^2$ as a
measure of fit quality.
For some negative ratios $r=x_0/v$ the real roots of
\eqref{eq:alpha_ratio_form} can disappear on a time interval, in which case
$\alpha_v(t)$ must become complex (or the solver can jump to a distinct real branch
near a pole such as $\alpha\approx 1$). In this work we report real-branch fits
whenever a stable real branch exists on $[t_1,t_2]$. For each parameter pair $(v,x_0)$ we solve \eqref{eq:alpha_linear_implicit} pointwise
in time on a grid $t\in[t_1,t_2]$ (in our computations $t_1=1$, $t_2=10$).
Because \eqref{eq:alpha_linear_implicit} can admit multiple branches, we track a
continuous branch by \emph{continuation} in $t$:
the root found at $t_k$ is used as the initial guess at $t_{k+1}$.
For the illustrative example $x(t)=t$ and $T_0=1$, this reduces to
\begin{equation}
\frac{\Gamma(2)}{\Gamma(2-\alpha_{v}(t))}\,\hat t^{\,1-\alpha_{v}(t)}=1,
\label{eq:alpha_linear_constraint}
\end{equation}
which we interpret in two anchoring scenarios:
\begin{itemize}
\item \textbf{Fixed anchor} at $\hat t=\tau_0$:
\begin{equation}
\frac{\Gamma(2)}{\Gamma(2-\alpha_{v}(t))}\,\tau_0^{1-\alpha_{v}(t)} = 1.
\label{gamma_fixed_simple}
\end{equation}

\item \textbf{Moving anchor}:
\begin{equation}
\frac{\Gamma(2)}{\Gamma(2-\alpha_{v}(t))}\,\hat t^{1-\alpha_{v}(t)} = 1.
\label{gamma_moving_simple}
\end{equation}
\end{itemize}

Clearly, the trivial case $\alpha_{v}(t) = 1$
 corresponds to standard Newtonian mechanics. However, when 
$\alpha_{v}(t) \neq 1$,
the representation deviates from the purely local description, introducing memory effects through fractional dimensions. This nontrivial solution offers an extended framework for understanding motion beyond the conventional integer-dimensional perspective. 
A numerical computation for moving anchor reveals a nontrivial solution for \( \alpha_{v}(t) \) exhibiting linear behavior as variable fractional order, which is well approximated by the equation:
\begin{equation}
\alpha_{v}(t) = (-2.63 \pm 0.05)t + (2.83 \pm 0.05). 
\label{alpha_fit}
\end{equation}
It can be observed that the trajectory follows a linear trend. However, the negative slope indicates a decreasing \( \alpha_{v}(t) \) within the time interval \( [1,10] \) seconds. This implies that the operator $\mathcal{O}^{\alpha_{v}(t)}$ is functioning as a fractional integral, incorporating non-local memory effects instead of acting as a conventional integer-order derivative. 
\noindent Equation \eqref{eq:alpha_asymptotic} explains the numerically observed
near-universal linear slope
\[
\alpha_v(t)\approx -2.63\,t
\]
on the finite interval $t\in[1,10]$.
Since $e\approx 2.718$, the numerical slope
$-2.62$ represents a pre-asymptotic approximation to the
true asymptotic value $-e$,
with subleading logarithmic corrections of order $O(\ln t)$.

\medskip
This example illustrates a representational duality: the same trajectory can be produced by a local integer-order description or by a nonlocal variable-order description. When Newtonian physics is chosen, with $\alpha_{v}(t) = 1$, the system is memoryless. However, when using a fractional-dimensional framework, where $\alpha_{v}(t) \approx (-2.63 \pm 0.05)t + (2.83 \pm 0.05)$, the system demonstrates memory effects. Remarkably, despite these differences, the observable trajectory
remains identical.
In the present formulation, the order trajectory $\alpha_{v}(t)$ emerges implicitly from constraint conditions imposed by the kinematics, rather than from a standalone evolution law. A central objective of future work is to obtain $\alpha_{v}(t)$ from first-principles physical calculations, instead of empirical fitting.
Table~\ref{tab:alpha_linear_examples} lists representative fitted lines
$\alpha_v(t)\approx m\,t+b$ obtained from \eqref{eq:alpha_linear_fit}.
These examples illustrate (i) the near-universal slope around $-2.62$ for the
stable decreasing branch and (ii) the sensitivity to sign-mismatched cases, where the
solver can converge to a near-constant branch $\alpha\approx 1$.

\begin{table}[H]
\centering
\caption{Linear fits for $\alpha_v(t)\approx m\,t+b$
solving \eqref{eq:alpha_linear_implicit}
(moving anchor, $t\in[1,10]$).}
\begin{tabular}{ccc}
\toprule
$(v,x_0)$ & Fitted $\alpha_v(t)$ & Regime \\
\midrule
$(1,0)$   & $-2.626585\,t+2.831281$ & stable decreasing \\
$(1,1)$   & $-2.610793\,t+1.347825$ & stable decreasing \\
$(1,5)$   & $-2.612108\,t-0.062139$ & stable decreasing \\
$(1,-1)$  & $-1.137440\,t+2.092730$ & deviation branch \\
$(1,-5)$  & $-0.308460\,t+1.958310$ & strong deviation \\
$(5,10)$  & $-2.607099\,t+0.741706$ & stable decreasing \\
$(-1,0)$   & $-2.626651\,t+2.831901$ & symmetric stable branch \\
$(-1,-1)$  & $-2.610898\,t+1.348768$ & symmetric stable branch \\
$(-10,5)$   & $0.000459\,t+0.996776$ & near-integer branch \\
$(-10,10)$  & $0.000000\,t+1.000000$ & exact integer branch \\
\bottomrule
\end{tabular}
\label{tab:alpha_linear_examples}
\end{table}

\subsection{Two-channel representation of constant acceleration}

A body subject to a constant net force undergoes motion with constant
\emph{physical} acceleration. In one spatial dimension, uniformly accelerated
motion is described by the Newtonian trajectory
\begin{equation}
x(t)=x(t_0)+v_0\,(t-t_0)+\tfrac12 a_0\,(t-t_0)^2,
\label{eq:xN_acc}
\end{equation}
where $v_0$ is the initial velocity and $a_0$ is the constant physical
acceleration.

Applying the two--channel decomposition \eqref{eq:two_channel_f} to $x(t)$, we
represent the same observable trajectory as
\begin{equation}
x(t)
=
x(t_0)
+
C_1\!\left(v_0\,\Delta t+\tfrac12 a_{\mathrm{loc}}\,\Delta t^2\right)
+
C_2\,B(t)\,\Delta t,
\qquad
\Delta t:=t-t_0,
\label{eq:traj_axiom2}
\end{equation}
where $C_1+C_2=1$, $C_1,C_2\ge0$, $a_{\mathrm{loc}}$ is the local-channel
acceleration parameter, and
\[
B(t):=\big(\mathcal{O}^{\alpha_{a}(t)}_{\mathrm{sc}}x\big)(\tau),
\qquad \tau\in\{t_0,t\},
\]
is the memory slope.

\begin{proposition}[Constant acceleration constraint on the memory slope]
\label{prop:acc_memory_slope}
If the physical acceleration is constant ($x''(t)=a_0$) and $C_2\neq 0$, then
the memory slope $B(t)$ must be affine in time.
\end{proposition}

\begin{proof}
Differentiating \eqref{eq:traj_axiom2} yields the velocity
\begin{equation}
v(t)
=
C_1\big(v_0+a_{\mathrm{loc}}\,\Delta t\big)
+
C_2\big(B(t)+\Delta t\,B'(t)\big),
\end{equation}
and the acceleration
\begin{equation}
a(t)
=
C_1 a_{\mathrm{loc}}
+
C_2\big(2B'(t)+\Delta t\,B''(t)\big).
\label{eq:acc_axiom2}
\end{equation}
Uniform acceleration requires $a(t)=a_0=\mathrm{const}$. For $C_2\neq0$, this
implies
\begin{equation}
2B'(t)+\Delta t\,B''(t)=b,
\qquad
b:=\frac{a_0-C_1 a_{\mathrm{loc}}}{C_2}.
\label{eq:B_ode_acc}
\end{equation}
The general solution is $B(t)=B_0+\frac{b}{2}\Delta t+\frac{K}{\Delta t}$.
The regularity at $t=t_0$ forces $K=0$, so the regular solution of \eqref{eq:B_ode_acc} is
\begin{equation}
B(t)=B_0+\tfrac{b}{2}\,\Delta t,
\end{equation}
which is affine in $\Delta t$.
\end{proof}
\begin{remark}[Affine memory slope and constant memory acceleration]
If $B(t)$ is affine,
\[
B(t)=B_0+\frac{b}{2}\Delta t,
\]
then $B''(t)=0$ and therefore the memory-channel contribution to acceleration is
constant:
\[
a_{\mathrm{mem}}(t)=2B'(t)+\Delta t\,B''(t)=2B'(t)=b.
\]
In particular, under the natural closure $a_{\mathrm{loc}}=a_0$ one has
$b=a_0$, hence $a_{\mathrm{mem}}=a_0$.
\end{remark}

\begin{corollary}[Determination of $B_0$ from the initial velocity]
Assume $B(t)$ is regular at $t=t_0$ and the representation reproduces the
Newtonian initial condition $v(t_0)=v_0$. Then
\begin{equation}
B_0=v_0.
\end{equation}
\end{corollary}

\begin{proof}
Setting $\Delta t=0$ in the velocity expression gives
$v(t_0)=C_1 v_0 + C_2 B_0$ (since $\Delta t\,B'(t)\to 0$ by regularity). Imposing
$v(t_0)=v_0$ and $C_1+C_2=1$ yields $B_0=v_0$ (for $C_2\neq 0$).
\end{proof}

\begin{remark}[Geometric-branch closure]
If one further chooses the natural closure $a_{\mathrm{loc}}=a_0$ (the local
channel carries the physical acceleration), then $b=a_0$ and the admissible
memory slope becomes
\begin{equation}
B(t)=v_0+\frac{a_0}{2}(t-t_0).
\end{equation}
This identification is a representational closure within the two-channel
framework and does not introduce new physical dynamics.

In the moving-anchor geometric branch ($\tau=t$), this yields the matching
condition
\begin{equation}
\big(\mathcal O^{\alpha_a(t)}_{\mathrm{sc}}x\big)(t)=v_0+\frac{a_0}{2}(t-t_0),
\end{equation}
which produces the nonlinear Gamma-based spectral equation for $\alpha_a(t)$ in
the next section.
\end{remark}

Uniform motion corresponds to a constant memory slope, while uniform
acceleration corresponds to an affine memory slope. Thus, constant physical
acceleration forces a linear evolution of $B(t)$; in this sense the nonlocal
channel carries kinematic curvature even when the observable trajectory
remains purely Newtonian.

Equation~\eqref{eq:acc_axiom2} shows that acceleration itself admits a
natural decomposition into instantaneous and memory--induced parts. Acceleration can therefore be written as
\begin{equation}
a(t)
=
C_1 a_{\mathrm{loc}}
+
C_2 a_{\mathrm{mem}},
\label{eq:acc_axiom22}
\end{equation}
where $a_{\mathrm{mem}}$ is equal to $  \big(2B'(t)+\Delta t\,B''(t)\big)$.
Even with fixed physical forcing, a time-dependent order $\alpha_a(t)$ can
redistribute kinematic weight between local and nonlocal channels. In this representational sense, the memory channel can redistribute
kinematic weight between channels without altering the physical
acceleration $x''(t)$.

\begin{theorem}[Local fractional-dimensional representation of a Newtonian trajectory]
\label{thm:main_representation}
Let $x(t)$ be a classical Newtonian trajectory satisfying
\[
m\,\ddot{x}(t)=F(x,t),
\]
on an interval $I\subset\mathbb{R}$, where $F$ is sufficiently smooth.
Fix an anchor $t_0\in I$, choose $\tau\in\{t_0,t\}$, and define a target
memory slope $S(t)$ appropriate to the kinematic level under consideration
(for example $S(t)=v(t)=\dot x(t)$ at the velocity level, or $S(t)=B(t)$
at the acceleration level).

Assume that the implicit constraint
\begin{equation}
\Phi(\alpha,t)
:=
\big(\mathcal{O}_{\mathrm{sc}}^{\alpha}x\big)(\tau)-S(t)=0
\label{eq:main_constraint}
\end{equation}
admits a point $(\alpha_*,t_*)$ such that
\begin{equation}
\Phi(\alpha_*,t_*)=0,
\qquad
\partial_\alpha \Phi(\alpha_*,t_*)\neq 0,
\label{eq:main_nondeg}
\end{equation}
and $\Phi$ is $C^1$ in a neighborhood of $(\alpha_*,t_*)$.

Then there exists an open interval $J\subset I$ containing $t_*$ and a unique
continuously differentiable order trajectory $\alpha(t)$ on $J$ such that
\begin{equation}
\big(\mathcal{O}_{\mathrm{sc}}^{\alpha(t)}x\big)(\tau)=S(t),
\qquad t\in J.
\label{eq:main_constraint_branch}
\end{equation}

Consequently, on $J$ the trajectory admits the two-channel representation
\begin{equation}
x(t)=C_1 x_{\mathrm{loc}}(t)+C_2 x_{\mathrm{mem}}(t),
\qquad
C_1,C_2\ge 0,\quad C_1+C_2=1,
\label{eq:main_two_channel}
\end{equation}
with
\begin{equation}
x_{\mathrm{mem}}(t)
=
x(t_0)
+
\big(\mathcal{O}_{\mathrm{sc}}^{\alpha(t)}x\big)(\tau)\,(t-t_0),
\label{eq:main_memory_repr}
\end{equation}
while the observable trajectory $x(t)$ remains exactly the original classical
Newtonian solution.
\end{theorem}

\begin{proof}
Fix $x(t)$ and the target slope $S(t)$. By assumption, the function
\[
\Phi(\alpha,t)=\big(\mathcal{O}_{\mathrm{sc}}^{\alpha}x\big)(\tau)-S(t)
\]
is $C^1$ in a neighborhood of $(\alpha_*,t_*)$, satisfies
$\Phi(\alpha_*,t_*)=0$, and is nondegenerate in the sense that
\[
\partial_\alpha \Phi(\alpha_*,t_*)\neq 0.
\]
Therefore, by the implicit function theorem, there exists an open interval
$J$ containing $t_*$ and a unique $C^1$ function $\alpha:J\to\mathbb{R}$ such
that
\[
\Phi(\alpha(t),t)=0
\qquad\text{for all }t\in J.
\]
This proves
\[
\big(\mathcal{O}_{\mathrm{sc}}^{\alpha(t)}x\big)(\tau)=S(t),
\]
which establishes the local existence of the order trajectory.

Now define the memory channel by
\[
x_{\mathrm{mem}}(t)
=
x(t_0)
+
\big(\mathcal{O}_{\mathrm{sc}}^{\alpha(t)}x\big)(\tau)\,(t-t_0),
\]
and let $x_{\mathrm{loc}}(t)$ denote the local channel chosen in the
two-channel representation. Since $C_1$ and $C_2$ are convex weights with
$C_1+C_2=1$, the decomposition
\[
x(t)=C_1x_{\mathrm{loc}}(t)+C_2x_{\mathrm{mem}}(t)
\]
is a representation of the same observable trajectory. No modification is made
to the Newtonian dynamics itself; the construction only introduces an internal
local/nonlocal decomposition of the already-given classical solution $x(t)$.
Hence the observable trajectory remains identical to the original Newtonian
trajectory.
\end{proof}

\section{Order Trajectory for Uniform Acceleration}
\label{sec:alpha_acceleration}

\subsection{Geometric-branch closure and the order equation}

Consider the uniformly accelerated Newtonian trajectory
\begin{equation}
x(t)=\frac{1}{2}a_{0} t^2 + v_{0} t + x_{0} ,
\label{eq:acc_traj}
\end{equation}
where $a_{0}$ is constant acceleration, $v_{0}$ the initial velocity, and $x_0$ the initial position. For simplicity we set the reference time $t_0=0$ in what follows.

In the two--channel fractional-dimensional framework with moving anchor $\tau=t$,
the order trajectory $\alpha_a(t)$ is determined only after choosing a closure.
On the \emph{moving-anchor geometric branch}, we close the representation by
identifying the memory slope with an affine target:
\begin{equation}
\boxed{
B(t)
:=\big(\mathcal{O}^{\alpha_a(t)}_{\mathrm{sc}}x\big)(t)
=\frac12\,a_{0}\,t+v_{0}.
}
\label{eq:axiom2_condition}
\end{equation}

Using the power-law rule for the Caputo/Riemann–Liouville unified operator,
we obtain

\begin{equation}
\boxed{
\frac{a_{0}}{\Gamma(3-\alpha)} t^{2-\alpha}
+
\frac{v_{0}}{\Gamma(2-\alpha)} t^{1-\alpha}
+
\frac{x_0}{\Gamma(1-\alpha)} t^{-\alpha}
= \frac12
a_{0} t + v_{0} .
}
\label{eq:gamma_acc_equation}
\end{equation}
Equation~\eqref{eq:gamma_acc_equation} is a nonlinear Gamma-based spectral equation for $\alpha_a(t)$.

\subsection{Dimensionless reduction for the geometric-branch equation}

Introduce the dimensionless variables
\begin{equation}
u = \frac{a_{0} t}{v_{0}}, 
\qquad
p = \frac{a_{0} x_0}{v_{0}^2},
\qquad
T = \frac{v_{0}}{a_{0}},
\qquad (a_{0}\neq 0,\ v_{0}\neq 0).
\label{eq:dim_vars}
\end{equation}
Then $t = T u$ and the right-hand side becomes
\[
\frac12 a_{0} t + v_{0}
=
\frac12 a_{0}(Tu)+v_{0}
=
\frac12 v_{0} u + v_{0}
=
v_{0}\!\left(1+\frac{u}{2}\right).
\]

Dividing \eqref{eq:gamma_acc_equation} by $v_{0}$ and substituting $t=Tu$ yields
\begin{equation}
\frac{a_{0}}{v_{0}}\frac{(Tu)^{2-\alpha}}{\Gamma(3-\alpha)}
+\frac{(Tu)^{1-\alpha}}{\Gamma(2-\alpha)}
+\frac{x_0}{v_{0}}\frac{(Tu)^{-\alpha}}{\Gamma(1-\alpha)}
=
1+\frac{u}{2}.
\end{equation}

Using $\frac{a_{0}}{v_{0}}=\frac{1}{T}$ and $\frac{x_0}{v_{0}}=pT$, each term becomes
\[
\frac{a_{0}}{v_{0}}(Tu)^{2-\alpha}
=
T^{1-\alpha}u^{2-\alpha},
\qquad
(Tu)^{1-\alpha}
=
T^{1-\alpha}u^{1-\alpha},
\qquad
\frac{x_0}{v_{0}}(Tu)^{-\alpha}
=
p\,T^{1-\alpha}u^{-\alpha}.
\]
Hence we obtain the compact dimensionless form
\begin{equation}
\boxed{
T^{1-\alpha}
\left[
\frac{u^{2-\alpha}}{\Gamma(3-\alpha)}
+
\frac{u^{1-\alpha}}{\Gamma(2-\alpha)}
+
p\,\frac{u^{-\alpha}}{\Gamma(1-\alpha)}
\right]
=
1+\frac{u}{2}.
}
\label{eq:dimensionless_acc_correct}
\end{equation}

\subsection{Asymptotic universal slope ($p=0$)}

To reveal the universal limit, set $p=0$ (i.e.\ $x_0=0$). For large $u$,
the dominant contribution inside the bracket is the first term
$\sim u^{2-\alpha}/\Gamma(3-\alpha)$, and the right-hand side satisfies
$1+\frac{u}{2}\sim \frac{u}{2}$. Equation \eqref{eq:dimensionless_acc_correct}
reduces to
\begin{equation}
T^{1-\alpha}\frac{u^{2-\alpha}}{\Gamma(3-\alpha)}
\sim
\frac{u}{2}.
\end{equation}
Dividing by $u$ and using $Tu=t$ gives
\begin{equation}
\boxed{
\frac{t^{1-\alpha}}{\Gamma(3-\alpha)}\sim \frac12.
}
\label{eq:asym_half}
\end{equation}

Let $\alpha=-\beta$ with $\beta>0$. Then \eqref{eq:asym_half} becomes
\begin{equation}
\boxed{
\frac{t^{1+\beta}}{\Gamma(\beta+3)}\sim \frac12.
}
\label{eq:beta_half}
\end{equation}
Taking logarithms yields
\begin{equation}
(1+\beta)\ln t - \ln\Gamma(\beta+3)\sim -\ln 2.
\label{eq:log_half}
\end{equation}
Using Stirling’s approximation
\[
\ln \Gamma(\beta+3)
=
\left(\beta+\tfrac52\right)\ln\beta
-\beta
+\tfrac12\ln(2\pi)
+O\!\left(\frac{\ln\beta}{\beta}\right),
\]
the dominant exponential balance implies
\begin{equation}
\boxed{
\beta(t)\sim e\,t,
\qquad
\alpha(t)\sim -e\,t,
\qquad (p=0,\ t\to\infty).
}
\label{eq:universal_et}
\end{equation}
Thus the Euler number  $e$ again governs the universal asymptotic slope.

\subsection{Next correction beyond the leading $-e t$ slope (geometric branch)}

To obtain the next correction, write
\begin{equation}
\beta(t)=\beta_0(t)+\delta(t),
\qquad
\beta_0(t)=e\,t,
\qquad
|\delta(t)|\ll\beta_0(t).
\end{equation}

We first note
\begin{equation}
\ln \beta_0 = \ln(et)=\ln t + 1.
\end{equation}

Expanding $\ln\beta$ around $\beta_0$ gives
\begin{equation}
\ln\beta
=
\ln\beta_0
+
\frac{\delta}{\beta_0}
+
O\!\left(\frac{\delta^2}{\beta_0^2}\right).
\end{equation}

Hence
\begin{align}
\ln t - \ln\beta + 1
&=
\ln t
-
\left(
\ln t + 1 + \frac{\delta}{\beta_0}
\right)
+1
\\
&=
-\frac{\delta}{\beta_0}.
\end{align}

The dominant nonlinear term therefore simplifies to
\begin{equation}
\beta(\ln t - \ln\beta + 1)
\approx
\beta_0\!\left(-\frac{\delta}{\beta_0}\right)
=
-\delta.
\end{equation}

Substituting into \eqref{eq:log_half} yields

\begin{equation}
-\delta
+
\left[
\ln t
-
\frac{5}{2}\ln\beta_0
-
\frac{1}{2}\ln(2\pi)
\right]
\approx
-\ln 2.
\end{equation}

Solving for $\delta(t)$ gives
\begin{equation}
\delta(t)
\approx
\ln t
-
\frac{5}{2}\ln\beta_0
-
\frac{1}{2}\ln(2\pi)
+
\ln 2.
\end{equation}

Using $\ln\beta_0=\ln t+1$, we obtain
\begin{align}
\delta(t)
&\approx
\ln t
-
\frac{5}{2}(\ln t+1)
-
\frac{1}{2}\ln(2\pi)
+
\ln 2
\\
&=
-\frac{3}{2}\ln t
-
\frac{5}{2}
-
\frac{1}{2}\ln(2\pi)
+
\ln 2.
\end{align}

Therefore the refined asymptotic expansion becomes
\begin{equation}
\boxed{
\beta(t)
\approx
e\,t
-
\frac{3}{2}\ln t
-
\frac{5}{2}
-
\frac{1}{2}\ln(2\pi)
+
\ln 2,
\qquad (p=0,\; t\to\infty).
}
\end{equation}

Recalling $\alpha(t)=-\beta(t)$, we finally obtain

\begin{equation}
\boxed{
\alpha(t)
\approx
- e\,t
+
\frac{3}{2}\ln t
+
\frac{5}{2}
+
\frac{1}{2}\ln(2\pi)
-
\ln 2.
}
\label{eq:stirling_correction} 
\end{equation}

We demonstrate this for $a_{0}=1$ using parameter pairs $(v_{0}, x_{0})$ from previous section as shown in tables 2 and 3. Table 2 quantifies the nonlinear transient regime, while Table 3 quantifies the asymptotic linear regime. We observe that the acceleration-level order trajectory exhibits two regimes:

\begin{enumerate}
\item \textbf{Early-time nonlinear regime:}  
Gamma nonlinearity dominates; branch competition and curvature are visible within the time interval $t\in[0,2]$.

\item \textbf{Late-time asymptotic regime:}  
The dominant branch becomes linear with universal slope $-e$ within the time interval $t\in[2,10]$.
\end{enumerate}
The logarithmic term of  Eq.~\eqref{eq:stirling_correction} explains why numerical linear fits over finite intervals (e.g.\ $t\in[2,10]$) yield slopes slightly smaller in magnitude than $-e$.
The constant $e$ arises from the exponential structure of the Gamma function under Stirling expansion, revealing a deep spectral structure in the fractional-dimensional kinematics.

\medskip

Remarkably, the same constant $e$ also governs the linear-motion case. This suggests that the Euler number   is a structural invariant of the Gamma-based order equation rather than a consequence of specific dynamics.

This result implies that within the fractional-dimensional framework, the order trajectory $\alpha_{a}(t)$ varies along the trajectory, capturing the nonlinear characteristics of acceleration in such spaces. 
At present, $\alpha_{a}(t)$ is inferred from the kinematic constraint; deriving
$\alpha_{a}(t)$ from first-principles physical calculations is left for future work.

\begin{table}[H]
\centering
\caption{Quadratic fits
$\alpha_a(t)\approx c_2 t^2+c_1 t+c_0$ on $t\in[0.1,2]$ for $a_{0}=1  (\frac{m}{s^2})$ and $a_{0}=10     (\frac{m}{s^2})$ ,
solving \eqref{eq:gamma_acc_equation}.}
\begin{tabular}{cc}
\toprule
$(v,x_0)$ & Fitted $\alpha_a(t)$ for $a_{0}=1$ on $t\in[0.1,2]$\\
\midrule
$(1,0)$   &$0.122154\,t^2-0.532771\,t+2.078910$ \\
$(1,1)$   & $-0.389222\,t^2-0.906222\,t+0.143679$  \\
$(1,5)$   & $-0.335594\,t^2-0.761827\,t+0.473640$\\
$(1,-1)$  & $0.000771\,t^2-0.280972\,t+2.073379$ \\
$(5,10)$  & $0.098281\,t^2-0.670115\,t+1.759449$\\
\bottomrule
\toprule
$(v,x_0)$ & Fitted $\alpha_a(t)$ for $a_{0}=10$ on $t\in[0.1,2]$ \\
\midrule
$(1,0)$   &$-0.324974\,t^2-0.571539\,t+0.891507$ \\
$(1,1)$   & $-0.567789\,t^2+0.006418\,t+0.343718$  \\
$(1,5)$   &  $-0.476331\,t^2-0.191154\,t-0.595249$\\
$(1,-1)$  &  $-0.305405\,t^2-0.523136\,t+0.996531$ \\
$(5,10)$  & $-0.467858\,t^2-0.535886\,t+0.131156$\\
\bottomrule
\end{tabular}
\label{tab:alpha_linear_examples1}
\end{table}

\begin{table}[H]
\centering
\caption{Linear fits
$\alpha_a(t)\approx c_1 t+c_0$ on $t\in[2,10]$ for $a_{0}=1 (\frac{m}{s^2})$ and $a_{0}=10 (\frac{m}{s^2})$ ,
solving \eqref{eq:gamma_acc_equation}.}
\begin{tabular}{cc}
\toprule
$(v,x_0)$ & Fitted $\alpha_a(t)$ for $a_{0}=1$  on $t\in[2,10]$\\
\midrule
$(1,0)$   & $-2.501692\,t+3.271209$ \\
$(1,1)$   & $-2.495800\,t+2.089540$\\
$(1,5)$   & $-2.486642\,t+0.565641$ \\
$(1,-1)$  & $-1.801225\,t+3.026746$ \\
$(5,10)$  & $-2.512327\,t+1.110508$\\
\bottomrule
\toprule
$(v,x_0)$ & Fitted $\alpha_a(t)$ for $a_{0}=10$ on $t\in[2,10]$ \\
\midrule
$(1,0)$   & $-2.449608\,t+3.803573$  \\
$(1,1)$   & $-2.461828\,t+3.496215$  \\
$(1,5)$   &  $-0.055914\,t+1.822257$ \\
$(1,-1)$  &  $-2.413399\,t+4.068961$\\
$(5,10)$  &$-0.033676\,t+1.614066$ \\
\bottomrule
\end{tabular}
\label{tab:alpha_linear_examples2}
\end{table}
\section{Harmonic oscillator with local and nonlocal acceleration}

Consider a one--dimensional mass--spring system with stiffness $k$ and mass $m$.
The classical equation of motion is
\begin{equation}
m\,\ddot x(t)+k\,x(t)=0,
\qquad 
\omega^2:=\frac{k}{m}.
\label{eq:sho_classical}
\end{equation}
Within the two--channel kinematic framework, the observable position
and acceleration are decomposed into local and memory contributions,
\begin{equation}
x(t)=C_1\,x_{\mathrm{loc}}(t)+C_2\,x_{\mathrm{mem}}(t),
\qquad
a(t)=C_1\,a_{\mathrm{loc}}(t)+C_2\,a_{\mathrm{mem}}(t).
\end{equation}
Substituting these expressions into Eq.~\eqref{eq:sho_classical} gives
\begin{equation}
C_1\!\left(m\,a_{\mathrm{loc}}(t)+k\,x_{\mathrm{loc}}(t)\right)
+
C_2\!\left(m\,a_{\mathrm{mem}}(t)+k\,x_{\mathrm{mem}}(t)\right)=0 .
\end{equation}
Since $C_1$ and $C_2$ represent independent weights of two kinematic channels
(with $C_1+C_2=1$ and $C_2\neq 0$ in general), the vanishing of the total force
implies that each channel must satisfy the same dynamical constraint. The local channel
therefore obeys the classical oscillator equation
\begin{equation}
m\,a_{\mathrm{loc}}(t)+k\,x_{\mathrm{loc}}(t)=0,
\end{equation}
while the memory channel must satisfy
\begin{equation}
m\,a_{\mathrm{mem}}(t)+k\,x_{\mathrm{mem}}(t)=0 .
\label{eq:mem_constraint}
\end{equation}
To describe the memory channel we introduce an affine representation
relative to a temporal anchor point $t_0$,
\begin{equation}
x_{\mathrm{mem}}(t)=B(t)\,\Delta t+x_0,
\qquad
\Delta t:=t-t_0 ,
\label{eq:xmem_definition}
\end{equation}
where $B(t)$ is a time--dependent memory slope and $x_0$ is the
reference position at the anchor time.
Differentiating Eq.~\eqref{eq:xmem_definition} with respect to time gives
\begin{equation}
\dot{x}_{\mathrm{mem}}(t)=B'(t)\,\Delta t+B(t),
\end{equation}
and therefore
\begin{equation}
a_{\mathrm{mem}}(t)
=\ddot{x}_{\mathrm{mem}}(t)
=2B'(t)+\Delta t\,B''(t).
\end{equation}
Substituting these expressions into Eq.~\eqref{eq:mem_constraint}
yields the differential equation for the memory slope
\begin{equation}
m\big(2B'(t)+\Delta t\,B''(t)\big)
+
k\big(B(t)\Delta t+x_0\big)=0 .
\label{eq:B_equation}
\end{equation}
Introduce the variable
\begin{equation}
y(t):=x_{\mathrm{mem}}(t)=B(t)\Delta t+x_0 .
\end{equation}
Since
\begin{equation}
y''(t)=2B'(t)+\Delta t\,B''(t),
\end{equation}
Eq.~\eqref{eq:B_equation} becomes
\begin{equation}
m\,y''(t)+k\,y(t)=0 .
\end{equation}

Therefore, the general solution is
\begin{equation}
x_{\mathrm{mem}}(t)
=
A_m\cos(\omega\Delta t)+B_m\sin(\omega\Delta t).
\end{equation}
At the anchor time $t=t_0$ we have $\Delta t=0$, giving
\begin{equation}
x_{\mathrm{mem}}(t_0)=A_m=x_0.
\end{equation}
The initial memory velocity is defined as
\begin{equation}
v_{m0} := \dot{x}_{\mathrm{mem}}(t_0).
\end{equation}
Differentiating the general solution yields
\begin{equation}
\dot{x}_{\mathrm{mem}}(t)
=
-\omega A_m\sin(\omega\Delta t)
+
\omega B_m\cos(\omega\Delta t).
\end{equation}
Evaluating at $t=t_0$ gives
\begin{equation}
v_{m0}=\omega B_m,
\end{equation}
so that
\begin{equation}
B_m=\frac{v_{m0}}{\omega}.
\end{equation}
The memory trajectory therefore becomes
\begin{equation}
x_{\mathrm{mem}}(t)
=
x_0\cos(\omega\Delta t)
+
\frac{v_{m0}}{\omega}\sin(\omega\Delta t).
\end{equation}
Using Eq.~\eqref{eq:xmem_definition}, the memory slope is
\begin{equation}
B(t)
=
\frac{x_0\big(\cos(\omega\Delta t)-1\big)
+
\frac{v_{m0}}{\omega}\sin(\omega\Delta t)}
{\Delta t}.
\end{equation}
The apparent singularity at $\Delta t=0$ is removable. Expanding the numerator
for small $\Delta t$ shows that
\[
B(t_0)=\lim_{\Delta t\to 0} B(t)=v_{m0}.
\]
Therefore the observable trajectory remains identical to the
classical harmonic oscillator solution, while the memory slope 
$B(t)$ provides an alternative representation of the oscillatory dynamics.
Thus the memory channel obeys the same harmonic oscillator dynamics
as the local channel and therefore does not introduce a new dynamical
law. Instead, it provides an alternative geometric representation of the
same harmonic motion in which the oscillatory trajectory is encoded
through the evolution of the memory slope $B(t)$.

\section{Conclusion}

In this work, we have developed a fractional--dimensional kinematic framework in
which memory is represented geometrically through a time-dependent fractional
order trajectory. By decomposing motion into local and nonlocal
kinematic channels, we demonstrated that classical Newtonian trajectories can be
reproduced exactly while admitting nontrivial internal memory structure.

A central result is that admissible order trajectories are not freely chosen
but are constrained by kinematic consistency. Uniform motion requires a
constant memory slope, whereas uniform acceleration requires a linearly
evolving memory slope. This establishes a geometric interpretation of
acceleration as curvature in the memory channel, even when the observable
trajectory remains purely Newtonian. The harmonic oscillator further illustrates
that familiar motions can be sustained by time-dependent fractional orders
without modifying the underlying forces or potentials.

Importantly, the order trajectory does not obey an independent
equation of motion in the present formulation. Instead, it emerges implicitly
from consistency conditions imposed at different kinematic levels. This
distinguishes the proposed framework from conventional fractional models and
highlights its role as a structural extension of classical mechanics rather
than a phenomenological modification.

The present work should be viewed as a foundational step. While it establishes
a coherent geometric and kinematic interpretation of fractional order, the
derivation of order trajectory from first-principles physical mechanisms remains an
open problem. Future directions include variational formulations with order
regularization, connections to differential geometry and effective mass
theories, and applications to systems where memory plays a dynamical role,
such as viscoelasticity, anomalous transport, and emergent inertial effects.

By reframing fractional order as a geometric degree of freedom rather than a
fixed parameter, this framework opens a new pathway toward understanding memory
as an intrinsic component of physical motion.\\
\\
\textbf{\Large{Part B: Outlook and Future Works, Scope and Interpretation}}\\

The framework developed in this work is representational rather than
dynamical in its present form. The observable trajectory $x(t)$ always
remains a solution of the classical Newtonian equations of motion.
No modification of forces, potentials, or conservation laws has been
introduced.

Instead, the contribution of the present formulation lies in the
demonstration that classical trajectories admit a consistent
two-channel fractional representation in which memory is encoded
geometrically through a variable-order operator.
The order trajectory $\alpha(t)$ is not postulated as an independent
degree of freedom; rather, it is determined implicitly by kinematic
consistency conditions.

It is important to emphasize that this construction does not predict
new observable deviations from Newtonian mechanics. The fractional
order does not modify the measured trajectory; it provides an
alternative internal description of that trajectory.
In this sense, the present framework differs fundamentally from
fractional dynamical models used to describe anomalous diffusion,
viscoelastic response, or nonlocal field theories
\cite{mainardi2010fractional,tarasov2011fractional}.

The universal asymptotic slope $-e$ obtained in both uniform motion
and uniform acceleration cases arises from the exponential structure
of the Gamma function under Stirling asymptotics.
It reflects the analytic structure of the implicit order equation
rather than a new physical constant or dynamical principle.

The primary open question is whether the order trajectory $\alpha(t)$
can be derived from a first-principles variational formulation,
geometric constraint, or microscopic mechanism.
Until such a derivation is established, $\alpha(t)$ should be
interpreted as a geometric descriptor emerging from consistency
requirements rather than as a new physical field.

Future developments may explore whether similar constructions can
lead to dynamical extensions in which the order trajectory interacts
with physical degrees of freedom. At present, however, the framework
should be viewed as a mathematically consistent geometric extension
of classical kinematics.

{\section{Axiomatic Frameworks for Geometric Memory}
The proposed model for studying physics from a fractional dimensional 
perspective is built on the following axioms:\\

\noindent\textbf{{Axiom 3: Wave Function Representation in Fractional Dimensions}}\\
A physical system can be described using a wave function $\Psi(S)$ that operates within fractional dimensions, not just the traditional integer dimensions. The parameter $S$, as defined in Equation \eqref{S}, is a vector consisting of independent variables such as time $(t)$ and spatial coordinates $(x_i, \,i=1,2,3 \ \text{or} \ \mathbf{r})$.
\begin{equation}
\begin{split}
S=[t, x_1, x_2, x_3].
\label{S}
\end{split}
\end{equation}
This suggests that fractional dimensions are not just mathematical abstractions but have real physical significance, allowing wave functions to encapsulate more complex behaviors and interactions than possible within the confines of integer dimensions.\\

\noindent\textbf{{Axiom 4:	Use of Normalized Fractional Derivatives}}\\
The calculation of wave functions within fractional dimensions should employ normalized fractional derivatives. These derivatives are specially adapted to the fractional nature of the dimensions they operate in, ensuring that the calculations reflect the unique properties of fractional spaces. This normalization is crucial to maintain consistency and accuracy in mathematical operations, enabling meaningful physical interpretations of the results.\\
\noindent The final wave function $\Psi$ can be obtained by applying a \textbf{normalized fractional derivative} to the initial wave function $\Psi_0$ along a trajectory in fractional dimensional space:
\begin{equation}
\begin{split}
\Psi =\mathcal{O}_{\rm N }^{\alpha(S)}\Psi_0,
\end{split}
\end{equation}
\noindent where $\mathcal{O}_{\rm N }$ is the normalized fractional operator, and $\alpha(S)$ is the fractional order associated with the trajectory of the system within the fractional dimensional space. \\

\noindent\textbf{{Axiom 5: Incorporation of Local and Nonlocal parts}}\\
A physical system is made up of both local (point-specific) and nonlocal (extended or entangled) parts. For any physical system, its wave function must reflect this duality. This axiom highlights the need for a holistic approach to wave function formulation, ensuring that both localized effects and more extended, nonlocal interactions are accurately represented, especially within fractional dimensional spaces where such dualities may manifest differently.
This relationship can also be expressed by considering the contributions of both local and nonlocal terms
\begin{equation}
\boxed{
\mathcal{O}_{\rm N }^{\alpha(S)}\Psi_0 = \text{Common term (Short Range Memory)} + \text{Extra term (Long Range Memory)}.}
\end{equation}
The behavior of nature can vary based on the specific conditions of a phenomenon. This variability is influenced by the nonlocal term, which is mathematically derived from the \textbf{nonlocal} components of the fractional derivative and depends on the specific definition of the fractional derivative. This axiom is also reflective of Gödel’s incompleteness theorem and also with some example in physics like entanglement as a nonlocal effect in QM or Butterfly effect [ref.].\\
Noted that the local part  provides an appropriate  approximation to the most problem in the classical physics.
\begin{equation}
\mathcal{O}_{\rm SRMN }^{\alpha(S)}\Psi_0 \approx \text{Short Range Memory},
\end{equation}
where the subscript ${\rm SRMN}$ denotes the short range memory (SRM) and normalized fractional  operator.\\

\noindent\textbf{{Axiom 6: Quantization of Dimensions Based on Photon Dimensions}}\\
Dimensions are quantized in a manner that is intrinsically linked to the dimensional characteristics of photons. Each dimension can be thought of as a discrete, quantized state, analogous to quantum states, which can be studied individually. This quantization implies that even within fractional dimensional spaces, there is a structured and orderly progression of dimensions, similar to the energy levels in quantum systems.
The photon is the only known particle with \textbf{the smallest dimension and the highest velocity}. We emphasize the importance of considering the photon's dimension, which can be defined using the wave-particle duality principle. The photon is modeled as a particle corresponding to a sphere with radius $R$, where the wavelength $\lambda$ corresponds to the perimeter of the sphere's enclosing circle ($\lambda=2n\pi R$ where n$\in \mathbb{N}$). Therefore, the photon's dimension can be described by the volume $V$ of this sphere as follows:
\begin{equation}
D_{\rm ph} = \frac{4\pi}{3}R^3 = \frac{1}{6n\pi^2}\lambda^3.
\end{equation}
\noindent From a fractional dimensional perspective, the photon's dimension varies between observers and can be determined through the Doppler frequency shift. The dimensional variation relative to the speed of light is given by:

\begin{equation}
\frac{dD_{\rm ph}}{d\nu} = -\frac{1}{2n\pi^2}\frac{c^3}{\nu^4}.
\end{equation}

It is observed that the dimension of the photon decreases with increasing frequency, leading to smaller dimensional variations at higher frequencies, approximately proportional to $\frac{1}{\nu^4}$. This effect is currently undetectable by existing optical devices. For instance, the dimensional variation of a photon at an 800 nm red light wavelength is $2.3\times10^{-35} \, \text{m}^3/\text{Hz}$. As the photon's dimension approaches zero, the speed of light remains constant for all observers, aligning with special relativity. This insight addresses the question of why the speed of light is absolute in special relativity: because the photon's dimension variation is effectively zero and it is invariant across all observers, regardless of space or time. However, from a fractional dimensional viewpoint, the photon has a dimension variation close to zero but not exactly zero.
The implications of axiom 4 lead to a novel form of quantization based on dimension, termed "quantized dimension". The photon's dimension ($D_{\rm ph}$) represents the smallest dimensional scale and serves as a unit for dimensional quantization.\\

\noindent\textbf{{Axiom 7: Observer-Dependent Dimensional Perception in Fractional Dimensional Space}}\\
The dimensions observed in any physical system are not absolute but depend on the observer's perspective, particularly the distance and angle relative to the object being observed in fractional dimensional space. This axiom introduces a relativistic element to dimensionality itself, suggesting that the dimensional nature of space could vary based on how it is measured or observed, adding a layer of complexity to physical observations and experiments.
\noindent Consider an observer in coordinate system $S$ and an object in system $S'$. The angle between these systems becomes acute as the distance increases, reaching 90 degrees when the distance approaches infinity (or when the object disappears from the observer's view). This rotation can be described using the fractional derivative (FDr) operator. As illustrated in Figure \ref{3}, an object in the $S'$ coordinate system appears smaller as it moves away from the observer in the $S$ system. This observation can be understood by considering the trajectory of the $S'$ coordinates, where $S'$ is not orthogonal to $S$ and forms a specific angle at each space-time position. As mathematically described in axiom 2, the variation in dimension follows a fractional derivative along a defined trajectory. Note that the scalar 1 in Figure \ref{3} results from the integer derivative of $x$ or $t$.
For clarity and to avoid confusion with visual perspective effects, we assume that all phenomena occur within fractional dimensional space.
\begin{figure}[h!]
    \centering
    \includegraphics[width=.7\linewidth]{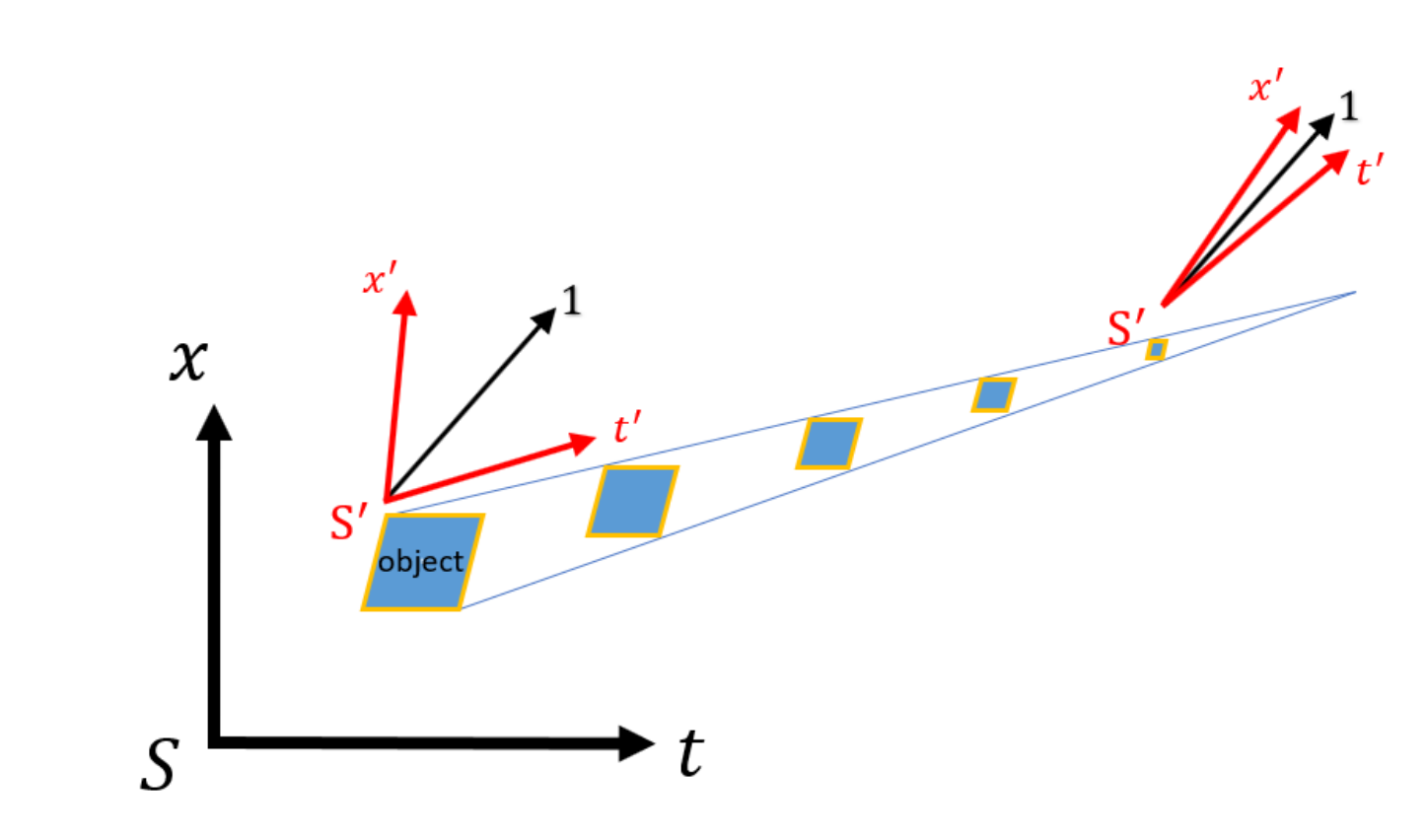}
    \caption{Schematic of $S$ and $S'$ coordinates in a movement framework of fractional dimensional space.}
    \label{3}
\end{figure}\\
\noindent \textbf{Hint:} For simplicity and as a preliminary investigation into the study of fractional dimensions in physics, we focus on the normalized local FDr component. The nonlocal part of the fractional derivative, which may explain non-locality effects in physics, remains under further investigation, and additional results will be presented in future work.

\subsection{\textbf{Normalized Local Fractional Derivatives of Key Functions}}

In this study, it is essential to determine the normalized local fractional derivatives (FDr) of trigonometric, exponential, and power series functions for further analysis. We will utilize the local term of the FDr, which is consistent across different definitions, or derive it directly by induction.
The following results are obtained by normalizing the fractional derivative (LFDr) through division by a normalization factor, ensuring the proper normalization of the basis functions. Detailed derivations can be found in the part B in the section 5 of the paper regarding mathematical framework.

\begin{equation}
\begin{split}
&D_{\rm N,L}^{\alpha}\left\{ \cos(\omega S)\right\} =\cos\left(\omega S+\frac{\pi}{2}\alpha\right),\\
&D_{\rm N,L}^{\alpha}\left\{ \sin(\omega S)\right\} = \sin\left(\omega S+\frac{\pi}{2}\alpha\right),\\&
D_{\rm N,L}^{\alpha}\left\{ e^{i\omega S}\right\}= e^{i\left(\omega S+\frac{\pi}{2}\alpha\right)},\\
&D_{\rm N,L}^{\alpha }\left\{ x^{n}\right\} 
=x^{(n-\alpha)},
\label{NLFD}
\end{split}
\end{equation}
\noindent where $D_{\rm N,L}^{\alpha}$ is the normalized local FDr operator.

\subsubsection{\textbf{Fractional Dimensional Basis Vector}}

Let $S$ and $S_{\alpha}$ be the basis vectors of the linear and fractional spaces, respectively. The relationship between these coordinates, as derived from Equation \eqref{NLFD}, is given by:

\begin{equation}
S_{\alpha} = D_{\rm N,L}^{\alpha}\left\{ S \right\} = S^{(1-\alpha)}.
\label{DS}
\end{equation}
In summary, the coordinate $S_{\alpha}$ (depicted by the red arrow in \figref{1}) for a specific value of $\alpha$ represents a rotation of the original coordinate $S$ within the fractional dimensional space. This rotation is achieved through the action of the LFDr operator with a corresponding order of $\alpha$.

\begin{figure}[h!]
    \centering
    \includegraphics[width=.40\linewidth]{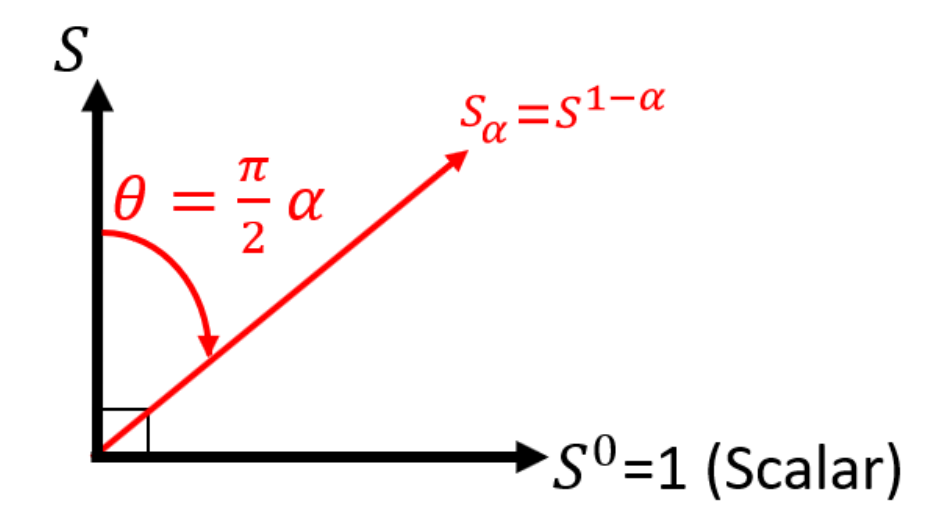}
    \caption{Illustration of the fractional dimensional basis vector $S_{\alpha}$ (red arrow) as a result of rotating the integer basis vector $S$ using the FDr operator. The vector rotates from 0 dimension to 1 ($0 \leq \alpha \leq 1$), equivalent to a rotation from 0 to 90 degrees ($0 \leq \theta \leq \frac{\pi }{2}$).}
    \label{1}
\end{figure}

\figref{2} demonstrates how a basic vector $x$, through rotation, can approach 1. The scalar value 1 is selected because the integer derivative of $x$ equals one. This outcome is achieved via a fractional derivative corresponding to the angle $\theta = \frac{\pi }{2}\alpha$. 

To represent the fractional space of space-time, we require three orthogonal axes: $x$ (representing $x_i$), $t$, and $1$, as depicted in \figref{2}(a). Here, $x$ and $t$ are vectors, while $1$ represents the scalar one. The red arrows in \figref{2}(b) to (d) indicate how the space-time axes behave as $\alpha$ approaches zero (\figref{2}(b)), with the angle $\theta$ near zero, and as $\alpha$ approaches one (\figref{2}(d)), with the angle $\theta$ approaching $\pi/2$.

\begin{figure}[h]
    \centering
    \includegraphics[width=1\linewidth]{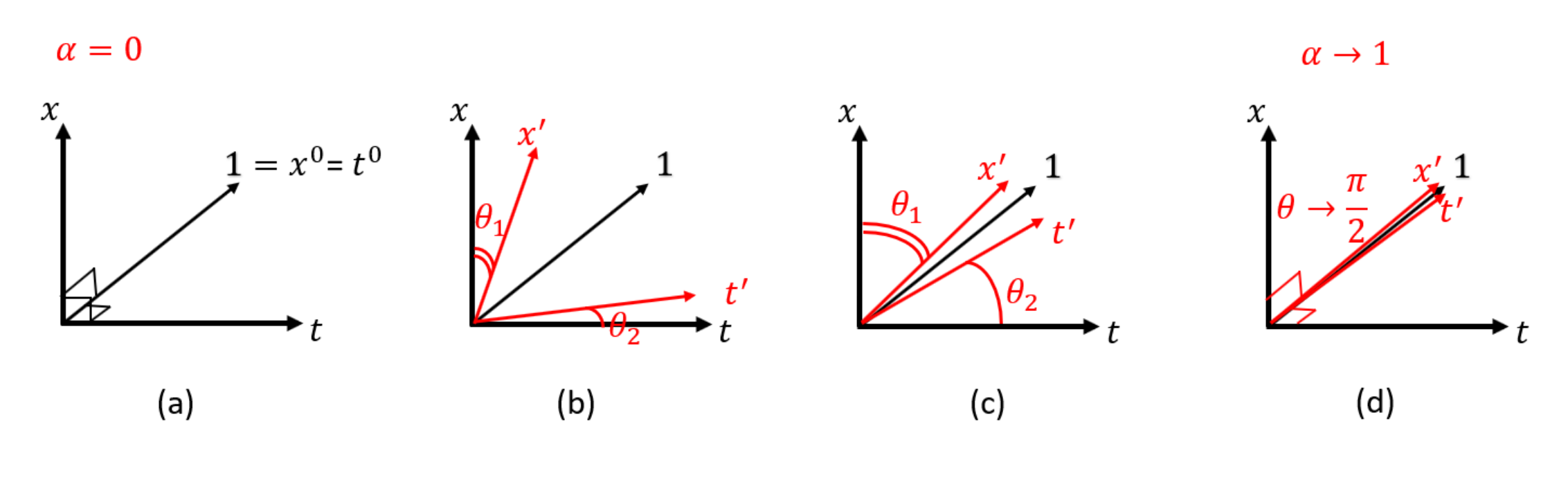}
    \caption{Rotation process of the vector axes $x$ and $t$ through fractional space. (a) All vector axes are orthogonal ($\theta=0$ or $\alpha=0$). As the system evolves, the vector axes $x$ and $t$ undergo rotation: (b) $\theta\approx0$ ($\alpha\approx0$), (c) $0<\theta<\pi/2$ ($0<\alpha<1$), and (d) $\theta\approx\pi/2$ ($\alpha\approx1$).}
    \label{2}
\end{figure}

As will be demonstrated in the subsequent section, various scenarios arise depending on the angle of rotation $\theta$.

\subsubsection{\textbf{Fractional Dimensional Basis Function}}

Let $\phi(S)$ and $\phi_{\alpha}(S)$ denote the basis functions of the linear and fractional spaces, respectively. The relationship between the basis functions in these spaces is expressed as:

\begin{equation}
\phi_{\alpha}(S) = \mathcal{O}_{\rm N}^{\alpha}\left\{ \phi(S) \right\}.
\label{FDS}
\end{equation}

Here, $\alpha$ represents a specific order of the fractional dimension for each space-time position.

\subsection{Space-Dimension-Time (SDT)}
The concept of space-time, rooted in relativity theory, suggests that space and time are coupled for a moving system. However, this coupling raises the question of why these seemingly independent quantities are interconnected. In our approach, we hypothesize that space and time are unified within a dimensional framework, which we term Space-Dimension-Time (SDT). This concept posits that space and time are coupled through a dimensional platform, as illustrated in Fig.~\ref{fig:SDT}. Although space and time are independent variables, they are interconnected by dimension through dynamical movement.

\begin{figure}[h!]
\centering
\includegraphics[width=.65\linewidth]{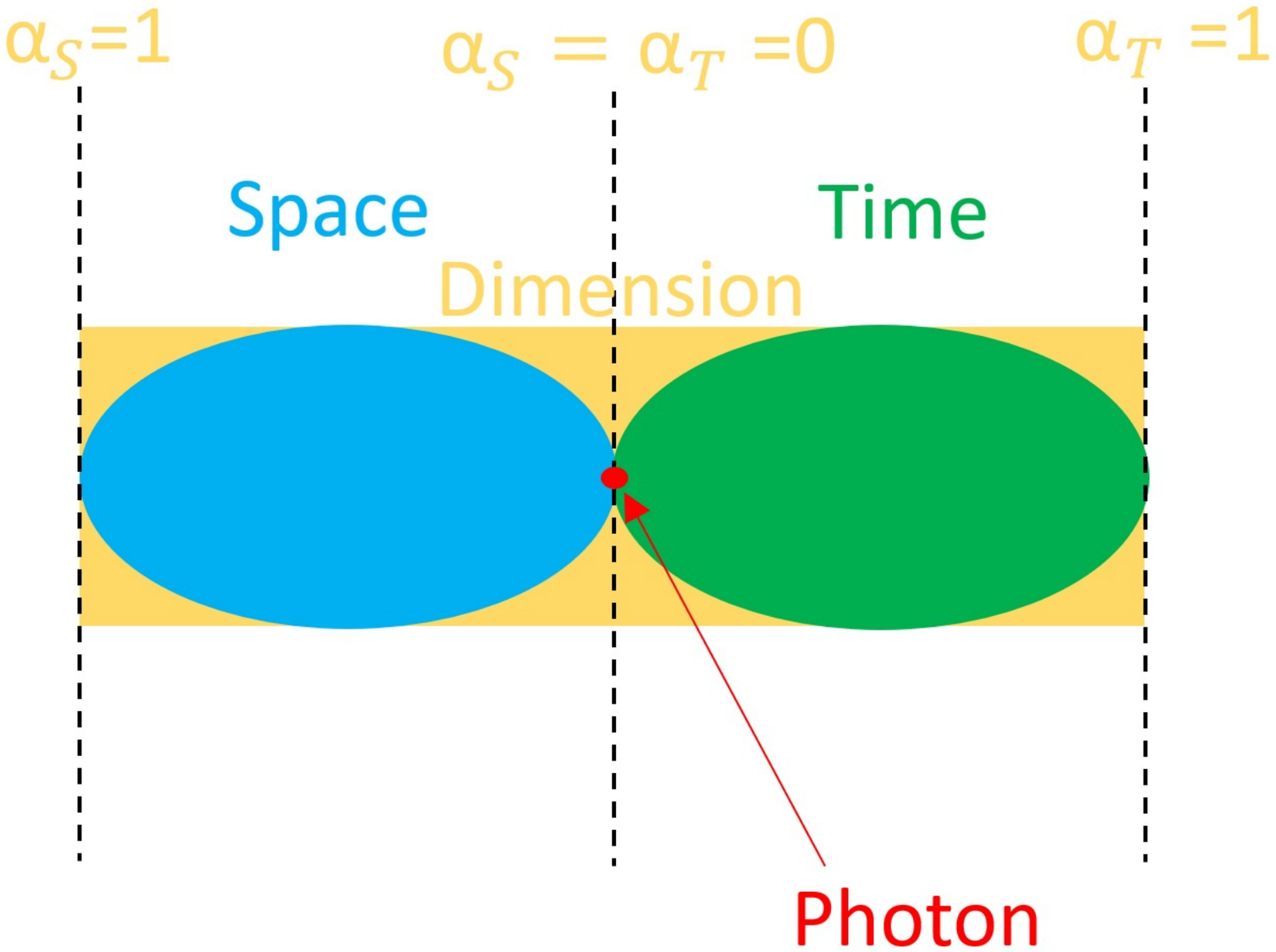}
\caption{The coupling of space and time through the dimension, termed Space-Dimension-Time (SDT). The dimension acts as a platform for space and time.}
\label{fig:SDT}
\end{figure}

In the static scenario, there is no coupling, as the system is steady. To demonstrate the coupling between space and time, one can consider both variables undergoing simple linear movement. Specifically, the $x_i$ and $t$ axes rotate towards the scalar 1-axis. Since space and time are independent, we express this relationship as follows:

\begin{equation}
\begin{split}
{\alpha}_i(x_i) &= \beta_i x_i + \widetilde{\alpha}_i \quad (i=1,2,3), \\
{\alpha}_4(t) &= \beta_4 t + \widetilde{\alpha}_4,
\end{split}
\label{eq:SDT_linear}
\end{equation}
where  $\widetilde{\alpha}_{i}$, $\beta_{i}$  are constant coefficients.
Taking the derivative of these expressions,

\begin{equation}
\begin{split}
d{\alpha}_i(x_i) &= \beta_i dx_i, \\
d{\alpha}_4(t) &= \beta_4 dt,
\end{split}
\end{equation}

and dividing them yields

\begin{equation}
\frac{d({\alpha}_i(x_i))}{d({\alpha}_4(t))} = \frac{\beta_i dx_i}{\beta_4 dt} = \frac{\beta_i}{\beta_4} v_i.
\label{eq:SDT_linear_relation}
\end{equation}

Equation \eqref{eq:SDT_linear_relation} indicates that for a constant velocity, the rotational movement of the $x_i$-axis relative to the $t$-axis is related to the velocity. In other words, from a dimensional perspective, space is related to time.\\

{\section{Toward a Dynamical Law for the Order Trajectory}}

In the preceding work, the fractional order trajectory $\alpha(t)$
emerged implicitly from kinematic consistency conditions.
Specifically, $\alpha(t)$ was determined so that the fractional-dimensional
memory slope reproduced a prescribed observable trajectory.
While this establishes a coherent representation theory,
$\alpha(t)$ remains a derived quantity rather than an independent
physical variable.

In this section we propose a minimal dynamical extension in which
$\alpha(t)$ is promoted to an internal degree of freedom whose evolution
is governed by a variational principle.
The aim is not yet to claim a new fundamental law, but to
formulate a candidate dynamical mechanism capable of selecting
$\alpha(t)$ from first principles.
\subsection{Extended action with a memory-geometry field}

Let $x(t)$ denote the observable coordinate and
$\alpha(t)$ the internal order trajectory.
We introduce the action functional
\begin{equation}
\boxed{
S[x,\alpha]
=
\int dt
\left[
\frac{m}{2}\dot x^2
-
V(x)
\right]
+
\frac{\eta}{2}\int dt\, \dot\alpha^2
+
\frac{\lambda}{2}\int dt\,
\Big(
\mathcal O^{\alpha(t)}_{sc} x - \dot x
\Big)^2 .
}
\label{eq:extended_action}
\end{equation}
The first term is the standard mechanical action.
The second term assigns kinetic cost to variations of the order trajectory,
penalizing rapid fluctuations of $\alpha(t)$ and introducing a smoothness
scale governed by $\eta>0$.
The third term couples the local velocity $\dot x$
to the fractional-dimensional memory slope
$\mathcal O^{\alpha}_{sc}x$,
with coupling strength $\lambda \ge 0$.
When $\lambda=0$, the theory reduces to classical mechanics.
When $\lambda>0$, the system dynamically penalizes mismatch
between instantaneous and memory-generated motion.

\subsection{Euler--Lagrange equations}

Variation with respect to $x(t)$ yields
\begin{equation}
m\ddot x + V'(x)
=
\lambda\,\frac{\delta}{\delta x}
\left(
\mathcal O^{\alpha}_{sc}x - \dot x
\right)
\left(
\mathcal O^{\alpha}_{sc}x - \dot x
\right),
\label{eq:x_equation}
\end{equation}

The right-hand side represents a memory-induced force arising
from the mismatch between the local and fractional-dimensional velocities.
In the limit $\lambda \to 0$, Newtonian dynamics is recovered.

Variation with respect to $\alpha(t)$ gives
\begin{equation}
\boxed{
\eta\,\ddot\alpha
=
\lambda
\Big(
\mathcal O^{\alpha}_{sc}x - \dot x
\Big)
\frac{\partial}{\partial \alpha}
\left(
\mathcal O^{\alpha}_{sc}x
\right).
}
\label{eq:alpha_equation}
\end{equation}
Equation~\eqref{eq:alpha_equation} governs the evolution of the
order trajectory.
The order field is driven by the mismatch between the
local and memory channels and relaxes according to the
geometric response factor
$\partial_\alpha(\mathcal O^{\alpha}_{sc}x)$.
Define the mismatch (memory-velocity error)
\begin{equation}
y(t):=\mathcal O^{\alpha(t)}_{sc}x(t)-\dot x(t).
\end{equation}
The coupling term in \eqref{eq:extended_action} is then $\frac{\lambda}{2}\int y^2\,dt$.
Holding $\alpha(t)$ fixed, we have
\[
\delta y=\mathcal O^{\alpha}_{sc}\delta x-\delta\dot x.
\]
Therefore
\begin{align}
\delta S
&=\int dt\Big(m\dot x\,\delta\dot x -V'(x)\delta x\Big)
+\lambda\int dt\,y\Big(\mathcal O^{\alpha}_{sc}\delta x-\delta\dot x\Big).
\end{align}
Integrating by parts the terms containing $\delta\dot x$ and introducing the adjoint
operator $(\mathcal O^{\alpha}_{sc})^\dagger$ defined by
\[
\int dt\,f\,(\mathcal O^{\alpha}_{sc}g)
=
\int dt\,\big((\mathcal O^{\alpha}_{sc})^\dagger f\big)\,g
\quad (\text{up to boundary terms}),
\]
we obtain
\begin{equation}
\delta S
=
\int dt\Big[
-m\ddot x -V'(x)+\lambda\big((\mathcal O^{\alpha}_{sc})^\dagger y+\dot y\big)
\Big]\delta x.
\end{equation}
Since $\delta x$ is arbitrary, the $x$--equation is
\begin{equation}
\boxed{
m\ddot x + V'(x)
=
\lambda\Big((\mathcal O^{\alpha}_{sc})^\dagger y+\dot y\Big),
\qquad
y=\mathcal O^{\alpha}_{sc}x-\dot x.
}
\label{eq:x_equation_correct}
\end{equation}
The right-hand side is a memory-induced force generated by the mismatch
between the local and fractional-dimensional velocities.
Holding $x(t)$ fixed, the only $\alpha$-dependence enters through
$\dot\alpha^2$ and $\mathcal O^{\alpha}_{sc}x$.
Since $\delta(\mathcal O^{\alpha}_{sc}x)=\partial_\alpha(\mathcal O^{\alpha}_{sc}x)\,\delta\alpha$,
we find
\begin{equation}
\delta S
=
\int dt\Big[
-\eta\ddot\alpha
+\lambda\,y\,\partial_\alpha(\mathcal O^{\alpha}_{sc}x)
\Big]\delta\alpha.
\end{equation}
Thus the order-trajectory equation is
\begin{equation}
\boxed{
\eta\,\ddot\alpha
=
\lambda
\Big(
\mathcal O^{\alpha}_{sc}x-\dot x
\Big)\,
\frac{\partial}{\partial \alpha}
\Big(
\mathcal O^{\alpha}_{sc}x
\Big).
}
\label{eq:alpha_equation}
\end{equation}

\subsection{Power--Law Behavior as a Manifestation of Fractional Dimensionality}


Power--law behavior thus emerges whenever the physical process
possesses a \emph{fractional dimension of evolution}:
space or time derivatives of noninteger order naturally generate
scale--free correlations.
The order function $\alpha(t)$ (or $\alpha(x,t)$ in extended systems)
quantifies this dimensional deviation and therefore provides
a unified explanation for the prevalence of algebraic scaling in
nature---from viscoelastic relaxation to critical phenomena.
In the integer limit $\alpha\!\to\!1$ the system recovers
exponential relaxation and finite correlation length,
while departures from this limit introduce the long tails
and scaling laws long associated with ``critical'' behavior.
Power--law responses are among the most universal signatures in physics.
They appear in critical phenomena, diffusion, dielectric relaxation,
viscoelasticity, turbulence, and even in socioeconomic or biological data.
Near continuous phase transitions, for instance, the Landau order parameter
($\psi$) follows $
 \psi \;\propto\; (T_c - T)^{\beta},$
while susceptibilities, correlation lengths, and specific heats exhibit
similar algebraic forms with critical exponents
$\gamma,\,\nu,\,\alpha$.
Although these exponents are measured with high precision, their
\emph{origin}---why the decay law becomes a power rather than exponential---remains conceptually open.
In the present framework this behavior arises naturally once the
differential order is allowed to be \emph{fractional}, that is,
when the underlying dynamics unfolds in a noninteger or
\emph{fractional--dimensional} space.

\subsection{From exponential to algebraic dynamics}

In classical (integer--order) kinetics,
a relaxation process obeying
\[
\frac{dx}{dt} = -\lambda x
\]
has the solution $x(t)=x_0 e^{-\lambda t}$:
the decay is exponential because differentiation is of integer order,
implying a single intrinsic timescale $1/\lambda$.

Replacing the derivative by a
\emph{fractional operator} of order $\alpha$,
\[
\mathcal{O}_t^{\alpha}x(t) = -\lambda x(t),
\qquad \alpha\in R,
\]
yields the Mittag--Leffler solution
\[
x(t) = x_0 E_{\alpha}\!\left(-\lambda t^{\alpha}\right)
\;\sim\; \frac{x_0}{\lambda\,\Gamma(1-\alpha)}\,t^{-\alpha}
\quad (t\!\to\!\infty).
\]
Thus a fractional order directly generates a power--law tail:
the system \emph{remembers} its past with a kernel
$K_\alpha(t)\propto t^{-\alpha}$.
This is the temporal analogue of spatial long--range correlations at
criticality.
Within the fractional--dimensional framework,
the observed exponent of the power law is not an accident but the
dynamical imprint of the effective dimension $\alpha$.

\section{{{Viscoelastic Memory and the Geometry of Temporal Response:
From Hereditary Dynamics to Structural Time}}}

Viscoelastic materials exhibit memory effects: their present mechanical response depends on the history of deformation \cite{BagleyTorvik}. In classical continuum mechanics, such memory is modeled phenomenologically through hereditary constitutive laws involving integral kernels or internal variables. Fractional calculus has emerged as an effective mathematical framework for describing viscoelastic memory, capturing experimentally observed power-law relaxation and creep with remarkable accuracy. In this work, we present a focused study of viscoelastic memory from both a mathematical and a physical perspective. We first review the historical development of viscoelastic constitutive modeling and the role of fractional operators. We then analyze the canonical stress-relaxation problem under step strain within a fractional viscoelastic framework. Finally, we offer a physical interpretation of viscoelastic memory that distinguishes between dynamical memory arising from coarse-graining and a deeper structural interpretation in which memory reflects the temporal organization of material response. The analysis clarifies the status of memory in classical mechanics and situates fractional viscoelasticity within a broader conceptual framework concerning the structure of time in physical response.
The foundations of linear viscoelasticity were laid in the early twentieth century through the work of Boltzmann, Volterra, and others. The central insight was that stress at time \(t\) depends on the entire strain history:
\begin{equation}
\sigma(t) = \int_0^t G(t-\tau)\,\dot{\varepsilon}(\tau)\,d\tau,
\end{equation}
where \(G(t)\) is the relaxation modulus. This hereditary formulation captures memory explicitly through convolution integrals.

While mathematically general, such models are phenomenological. The kernel \(G(t)\) is fitted to experimental data, and its physical interpretation is indirect. Memory is treated as a consequence of unresolved microstructural processes rather than as a fundamental aspect of temporal response.

\subsubsection{Fractional calculus in viscoelasticity}

Fractional calculus emerged as a powerful tool for modeling viscoelasticity when it was recognized that many materials exhibit relaxation moduli with power-law decay. Fractional derivatives interpolate naturally between elastic and viscous behavior and require fewer parameters than classical multi-element models.

A basic fractional constitutive relation takes the form
\begin{equation}
\sigma(t) = \eta\, D_t^\alpha \varepsilon(t), \qquad 0<\alpha<1,
\end{equation}
where \(D_t^\alpha\) denotes a fractional derivative, typically of Caputo type. More realistic models, such as the fractional Zener model, combine elastic and fractional elements to ensure finite equilibrium response.

Despite extensive experimental validation, fractional viscoelasticity is often interpreted instrumentally, as an effective description rather than as a reflection of temporal structure.

\subsection{Mathematical Framework: Stress Relaxation in Fractional Viscoelasticity}
We consider the classical stress-relaxation experiment. A material is subjected to a sudden strain at time \(t=0\),
\begin{equation}
\varepsilon(t) = \varepsilon_0 H(t),
\end{equation}
where \(H(t)\) is the Heaviside step function. The strain is then held constant, and the stress response \(\sigma(t)\) is measured for \(t>0\).
A widely used linear viscoelastic model capable of describing stress relaxation is the fractional Zener model:
\begin{equation}
\sigma(t) + \tau_\sigma^\alpha D_t^\alpha \sigma(t)
=
E\left[ \varepsilon(t) + \tau_\varepsilon^\alpha D_t^\alpha \varepsilon(t) \right],
\end{equation}
where \(E\) is a modulus, \(\tau_\sigma\) and \(\tau_\varepsilon\) are characteristic times, and \(0<\alpha<1\).

Taking the Laplace transform and solving for the stress under step strain yields the relaxation modulus
\begin{equation}
G(t) = E_\infty + (E_0 - E_\infty)
E_\alpha\!\left[-\left(\frac{t}{\tau}\right)^\alpha\right],
\end{equation}
where \(E_\alpha(\cdot)\) is the Mittag--Leffler function, \(E_0\) is the instantaneous modulus, and \(E_\infty\) is the long-time modulus.

The stress response is therefore
\begin{equation}
\sigma(t) = \varepsilon_0 G(t).
\end{equation}

\subsection{Memory signature}

Unlike classical exponential relaxation, the Mittag--Leffler function exhibits a power-law tail at long times:
\begin{equation}
\sigma(t) - \varepsilon_0 E_\infty \sim t^{-\alpha}.
\end{equation}
This slow decay reflects persistent memory of the initial deformation, a hallmark of viscoelastic behavior.

\subsection{ Dynamical versus Structural Memory}

In conventional interpretations, viscoelastic memory is understood as dynamical. The material is assumed to possess many internal degrees of freedom—polymer chains, molecular rearrangements, microstructural relaxation modes—that are not explicitly modeled. Coarse-graining over these degrees of freedom leads to effective equations in which stress depends on strain history.

In this view, memory is not fundamental. It is an emergent consequence of incomplete description and would disappear if all microscopic variables were retained.

An alternative perspective is suggested by the structure of fractional constitutive laws themselves. Fractional derivatives do not add interaction terms or hidden variables; they modify the meaning of temporal differentiation. The present response is defined through an integral over a finite temporal domain, implying that the present is not an instantaneous point but a temporally extended region.

From this standpoint, memory is not a force or interaction but a geometric property of temporal response. The past does not influence the present causally; rather, the present structurally contains the past through the organization of time in the constitutive law.

This interpretation reframes viscoelastic memory as a manifestation of nontrivial temporal structure within classical mechanics. Integer-order models correspond to a limiting case in which the present collapses to an instant and memory disappears. Fractional models generalize this picture by allowing the temporal geometry of response to acquire structure.

While this does not negate the role of microstructure or coarse-graining, it suggests that viscoelastic memory may be understood at a deeper level as reflecting how material response unfolds in time.

The viscoelastic stress-relaxation problem illustrates how memory enters classical mechanics in a controlled and experimentally verifiable way. Fractional viscoelastic models provide not only accurate fits to data but also a window into the temporal organization of material response.

Whether this temporal structure should be regarded as fundamental or merely effective remains an open question. Nevertheless, the analysis shows that classical mechanics already contains phenomena that challenge the notion of an instantaneous present and motivate broader reflection on the structure of time in physical response.

In summary, we want to present a focused analysis of viscoelastic memory through the canonical stress-relaxation problem. Using fractional constitutive models, we demonstrated how memory manifests mathematically and physically in classical mechanics. Beyond their phenomenological success, fractional models invite a structural interpretation in which memory reflects the temporal geometry of response rather than a dynamical interaction.

This perspective situates viscoelasticity within a broader conceptual framework concerning the nature of time, memory, and evolution in physical systems, and suggests that even within classical mechanics, the structure of time may be richer than traditionally assumed.

{\section{Fractional-Dimensional Field Representation and Memory Structure}}

\subsection{Two-Channel Field Decomposition}

We begin from the fundamental assumption that the observable
electromagnetic field admits a two-channel representation,

\begin{equation}
\mathbf{E}_{\mathrm{obs}}
=
\mathbf{E}_{\mathrm{loc}}
+
\mathbf{E}_{\mathrm{mem}},
\qquad
\mathbf{B}_{\mathrm{obs}}
=
\mathbf{B}_{\mathrm{loc}}
+
\mathbf{B}_{\mathrm{mem}} .
\label{eq:field_decomposition}
\end{equation}

The local channel is defined as the classical Maxwell field satisfying
the standard vacuum equations without sources $\rho$ and $ J$,

\begin{equation}
\nabla \cdot \mathbf{E}_{\mathrm{loc}} = 0,
\qquad
\nabla \cdot \mathbf{B}_{\mathrm{loc}} = 0,
\end{equation}

\begin{equation}
\nabla \times \mathbf{E}_{\mathrm{loc}}
=
- \partial_t \mathbf{B}_{\mathrm{loc}},
\qquad
\nabla \times \mathbf{B}_{\mathrm{loc}}
=
\mu_0 \varepsilon_0 \partial_t \mathbf{E}_{\mathrm{loc}}.
\label{eq:maxwell_local}
\end{equation}

Thus the local sector preserves the integer-order differential structure
of classical electrodynamics.

The memory channel does not modify the Maxwell operators.
Instead, it provides an alternative fractional-dimensional representation
of the same observable field, governed by a spacetime-dependent order
field $\alpha(x,t)$.

\subsection{{A Signal Representation of System via Fractional Dimensional Basis Function}}

In this section, we demonstrate the signal behavior based on three scenarios. By using the assumption \eqref{FDS}, the formulation between the basis function of the transmitter ($T(S)$) and receiver ($R(S)$) signals can be rewritten as follows:
\begin{equation}
R(S) = \frac{1}{2}D_{\rm N}^{\alpha }\big\{ T(S) \big\} + \frac{1}{2}\bigg(D_{\rm N}^{\alpha }\big\{ T(S) \big\} \bigg)^* = \Re \bigg\{ D_{\rm N}^{\alpha }\big\{ T(S) \big\} \bigg\},
\label{RT}
\end{equation}
where $\Re$ and $*$ denote the real part of the basis function and complex conjugate, respectively.\\
The FDr operator connects the input signal (initial signal) to the output signal of the system through a trajectory in fractional dimension. This assumption denotes that while the transmitter itself is in an integer coordinate, from the receiver's point of view, it resides in a fractional dimensional coordinate. \\
For simplicity, we use two variables: $x_1$ (space) and $t$ (time), where $x_1$ is symbolized by $z$. Henceforth, $z$ and $t$ denote space and time variables, respectively.

\subsubsection{\textbf{Static Scenario}}

From a signal perspective, a constant distance is maintained between the transmitter and receiver as shown in Fig.~\ref{5}(a). This is analogous to an observer viewing a stationary object from a fixed distance.
\begin{figure}[h!]
	\centering
	\includegraphics[width=1\linewidth]{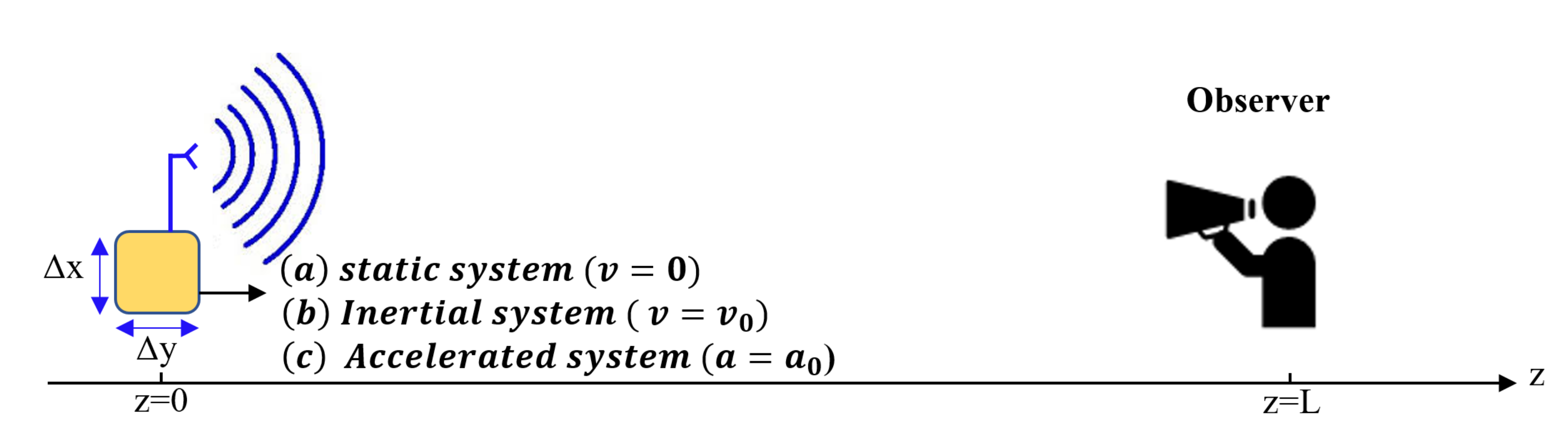}
	\caption{Schematic of three scenarios (a) static ($v=0$), (b) linear dynamic ($v=v_0$), (c) accelerated system ($a=a_0$).}
	\label{5}
\end{figure}

Consider a signal of the form $A_0\cos(\omega t + k z)$ propagating in free space. How does the received signal diagram change $(R(z=L,t))$ after propagating a distance $L$ in (a) a lossless and (b) a lossy medium? Here, $A_0$, $\omega$, and $k$ represent the amplitude, angular frequency, and wave vector, respectively.

**Hint:** When an electromagnetic wave propagates in a lossy medium, the wave vector is a complex value as:
\begin{equation}
k = \beta + i\alpha_{\text{att.}},
\end{equation}
where $\beta$ and $\alpha_{\text{att.}}$ are the propagation constant and attenuation decay coefficient, respectively.

The answer for part (a) is as follows:

From an electrodynamics standpoint, we know that the signal undergoes a phase shift corresponding to $kL$, where in a lossless medium, $k$ equals the propagation constant $\beta$:
\begin{equation}
R(z=L,t) = A_0\cos(\omega t + \beta L) = A_0\cos(\omega t + \Phi),
\end{equation}
where $\Phi = \beta L$.

However, from a fractional dimensional perspective, the received signal at $z=L$ is obtained by a constant fractional dimension of $\alpha_0$. Therefore, we have:
\begin{equation}
R(z=L,t) = D_{\rm N,L}^{\alpha_0 }\big\{ A_0\cos(\omega t) \big\} = A_0\cos(\omega t + \frac{\pi \alpha_0}{2}) = A_0\cos(\omega t + \Phi),
\end{equation}
where $\Phi = \beta L = \frac{\pi \alpha_0}{2}$ leading to realization of $\alpha_0= \frac{2}{\pi}\beta L$.

The answer for part (b) is as follows:

From an electrodynamics perspective, the final received signal after simplification at distance $L$ is:
\begin{equation}
R(z=L,t) = A_0e^{-\alpha_{\text{att.}} L}\cos(\omega t + \beta L),
\end{equation}
and from a fractional dimensional perspective, we have:
\begin{equation}
R(z=L,t) = \frac{1}{2}D_{\rm N,L}^{\alpha_0 }\big\{ e^{i\omega t} \big\} + \frac{1}{2}\bigg(D_{\rm N,L}^{\alpha_0 }\big\{ e^{i\omega t} \big\} \bigg)^* = \frac{1}{2}D_{\rm N,L}^{\alpha_0 }\big\{ e^{i\omega t} \big\} + \frac{1}{2}\bigg(D_{\rm N,L}^{\alpha_0^* }\big\{ e^{-i\omega t} \big\} \bigg),
\end{equation}
where $\alpha_0$ has a complex value as follows:
\begin{equation}
\alpha_0 = \alpha_0^{\text{R}} + i\alpha_0^{\text{I}}.
\end{equation}

Then, the complex fractional basis function is:
\begin{equation}
D_{\rm N,L}^{\alpha_0 }\big\{ e^{i\omega t} \big\} = e^{i(\omega t + \frac{\pi \alpha_0}{2})} = e^{i(\omega t + \frac{\pi}{2}\alpha_0^{\text{R}} + i\frac{\pi}{2}\alpha_0^{\text{I}})} = e^{-\frac{\pi}{2}\alpha_0^{\text{I}}}e^{i(\omega t + \frac{\pi}{2}\alpha_0^{\text{R}})},
\end{equation}
where $D_{\rm N,L}^{\alpha_0^* }\big\{ e^{-i\omega t} \big\}$ is the complex conjugate of $D_{\rm N,L}^{\alpha_0 }\big\{ e^{i\omega t} \big\}$.

Thus, after rearrangement, we have:
\begin{equation}
R(z=L,t) = A_0 \Re\bigg\{ D_{\rm N,L}^{\alpha_0 }(\big\{ e^{i\omega t} \big\}) \bigg\} = A_0e^{-\frac{\pi}{2}\alpha_0^{\text{I}}}\cos(\omega t + \frac{\pi}{2}\alpha_0^{\text{R}}),
\end{equation}

It can be observed that signal decay in a linear vector space corresponds to a trajectory of the signal in the imaginary domain of fractional dimension. Therefore, we have:
\begin{equation}
\frac{\pi}{2}\alpha_0^{\text{R}} = \beta L,
\end{equation}
and
\begin{equation}
\frac{\pi}{2}\alpha_0^{\text{I}} = \alpha_{\text{att.}} L.
\end{equation}

\subsubsection{\textbf{Dynamic Scenario}}

In this case, at least one of the transmitter or receiver is moving in fractional dimensions, which leads to what we call dynamic observation. Two types of movements, linear and quadratic trajectories, are demonstrated.

\paragraph{\textbf{Linear Dynamics}}
We demonstrate that two important cases in physics, the wave equation and the Doppler effect, can be understood as equivalent to a system moving in a fractional-dimensional space with a linear trajectory.
\subparagraph{Equivalence in the Wave Equation}
We begin by reformulating the wave equation based on the proposed assumption of fractional-dimensional space. As is well-known, a wave is a phenomenon that propagates in both time and space. This propagation can be modeled as a temporal (or spatial) signal moving through a spatial (or temporal) domain along a fractional-dimensional trajectory, denoted by $\alpha(\mathbf{r})$ or $\alpha(t)$. The wave function $\mathbf{\Psi}(\mathbf{r}, t)$ can be decomposed into two independent functions of time $t$ and space $\mathbf{r}$, such that $\mathbf{\Psi}(\mathbf{r}, t) = \psi_1(t)\psi_2(\mathbf{r})$. Therefore, the wave equation can be expressed as follows:
\begin{equation}
D_{\rm N}^{\alpha_1(\mathbf{r}) }\left\{\psi_1(t)\right\} = D_{\rm N}^{\alpha_2(t)}\left\{\psi_2(\mathbf{r})\right\}.
\label{wave}
\end{equation}
The left-hand side of equation \eqref{wave} implies that the temporal signal $\psi_1(t)$ moves through the spatial domain along a trajectory $\alpha_1(\mathbf{r})$, and must equal the right-hand side, which describes the spatial signal $\psi_2(\mathbf{r})$ moving through the time domain along a trajectory $\alpha_2(t)$.

For simplicity, as known from quantum mechanics and electrodynamics, the wave basis function for two variables, $z$ and $t$, is given by $\mathbf{\Psi}(z,t) = e^{i(\omega t + kz)}$. Thus, the fractional-dimensional wave equation becomes:
\begin{equation}
D_{\rm N,L}^{\alpha_1(z)}\left\{e^{i\omega t}\right\} = D_{\rm N,L}^{\alpha_2(t)}\left\{e^{ikz}\right\} = e^{i\left(\omega t + \frac{\pi}{2}\alpha_1(z)\right)} = e^{i\left(kz + \frac{\pi}{2}\alpha_2(t)\right)} = e^{i(\omega t + kz)},
\end{equation}
where the trajectories $\alpha_1(z)$ and $\alpha_2(t)$ are defined as:
\begin{equation}
\begin{split}
\alpha_1(z) &= \frac{2kz}{\pi}, \\
\alpha_2(t) &= \frac{2\omega t}{\pi}.
\end{split}
\end{equation}

\subparagraph{Doppler Effect}
As illustrated in Fig. 5(b), for an inertial system where an observer remains at a fixed position (e.g., $z=L$) and the transmitter moves towards (or away from) the receiver with velocity $v_t$, this phenomenon is known as the Doppler effect. Here, we demonstrate how the movement of the signal source along a linear trajectory in a fractional dimension is equivalent to the Doppler effect.

Consider a signal $T(z=0, t)$ at the transmitter with the expression:
\begin{equation}
T(z=0,t) = A_{0}\cos(\omega t).
\label{RT}
\end{equation}
Given the trajectory:
\begin{equation}
\alpha(z, t) = \beta_{41}z + \beta_{44}t,
\label{35}
\end{equation}
where $\beta_{41}$ and $\beta_{44}$ are constants, and setting $z$ as a constant value with respect to $t$, the received signal becomes:
\begin{equation}
R(z=L, t) = D_{\rm N,L}^{\alpha(S)}\left\{A_{0}\cos(\omega t)\right\} = A_{0}\cos\left(\omega t + \frac{\pi\alpha(S)}{2}\right) = A_{0}\cos\left(\omega t + \frac{\pi(\beta_{41} z + \beta_{44}t)}{2}\right).
\label{21}
\end{equation}
Rearranging gives:
\begin{equation}
R(z=L, t) = A_{0}\cos\left(\omega_D t + \Phi\right),
\end{equation}
where the constant phase $\Phi$ is:
\begin{equation}
\Phi = \frac{\pi}{2}\beta_{41}L,
\end{equation}
and $\omega_D$ is the new (Doppler) angular frequency:
\begin{equation}
\omega_D = \omega + \frac{\pi\beta_{44}}{2} = 2\pi(f + \Delta f) = 2\pi f_D,
\end{equation}
with the Doppler frequency shift \cite{eddington1926einstein}:
\begin{equation}
f_D = \left(\frac{c \pm v_r}{c \pm v_t}\right)f = f + \left(\frac{\pm v_r \mp v_t}{c \pm v_t}\right)f = f + \Delta f,
\end{equation}
where $c$ is the propagation speed of the wave in the medium and $v_t$ (corresponds to $v_0$ in \figref{5})  and $v_r$ are transmitter and receiver velocity, respectively. If the receiver moves towards (or away from) the source, the sign is positive (or negative). For this scenario, the observer velocity $v_r$ is zero ($v_r = 0$), and $\Delta f$ is:
\begin{equation}
\Delta f = \left(\frac{\mp v_t}{c \pm v_t}\right)f = \frac{\beta_{44}}{4}.
\end{equation}

\subsubsection{\textbf{Acceleration Dynamics Scenario}}
In this case, the trajectory in the fractional dimension is a quadratic function. We assume the following trajectory:
\begin{equation}
\alpha(S) = \alpha(z, t) = \gamma_{41}z^2 + \gamma_{44}t^2 + \beta_{41}z + \beta_{44}t,
\end{equation}
where $\beta_{41}$, $\beta_{44}$, $\gamma_{41}$, and $\gamma_{44}$ are constants. If the $\gamma$ coefficients are zeroed, the Doppler problem is retrieved. Therefore, the basis function for an accelerated system can be represented as:
\begin{equation}
\phi^{\alpha(z, t)}(S) = D_{\rm N}^{\gamma_{41}z^2 + \gamma_{44}t^2 + \beta_{41}z + \beta_{44}t}\left\{\phi(S)\right\},
\end{equation}
and the received signal for constant $z$ versus $t$ is:
\begin{equation}
\begin{split}
R(z, t) &= D_{\rm N,L}^{\alpha(z, t)}\left\{A_{0}\cos(\omega t)\right\} = A_{0}\cos\left(\omega t + \frac{\pi\alpha(z, t)}{2}\right)\\
&= A_{0}\cos\left(\omega t + \frac{\pi(\gamma_{41}z^2 + \gamma_{44}t^2 + \beta_{41}z + \beta_{44}t)}{2}\right).
\end{split}
\end{equation}
Rearranging gives:
\begin{equation}
R(z=L, t) = A_{0}\cos\left(\omega_{Da}(t)t + \Phi\right),
\end{equation}
where the constant phase at $z = L$ is:
\begin{equation}
\Phi = \frac{\pi}{2}(\gamma_{41}L^2 + \beta_{41}L),
\end{equation}
and $\omega_{Da}(t)$ is a new time-dependent (Doppler) angular frequency given by:
\begin{equation}
\omega_{Da}(t) = \frac{\pi\gamma_{44}}{2}t + \omega + \frac{\pi\beta_{44}}{2} = \frac{\pi\gamma_{44}}{2}t + \omega_D.
\label{WD}
\end{equation}
This corresponds to the gravitational red (or blue) shift. As a simple example, the profile of \eqref{WD} with all coefficients normalized to one is depicted in \figref{6}. It can be observed that this profile exhibits behavior consistent with gravitational red (or blue) shift \cite{eddington1926einstein}. Without considering the envelope, it also resembles the behavior of gravitational waves \cite{anderson2012gravitational}.\\
\begin{figure}[H]
	\centering
	\includegraphics[width=1\linewidth]{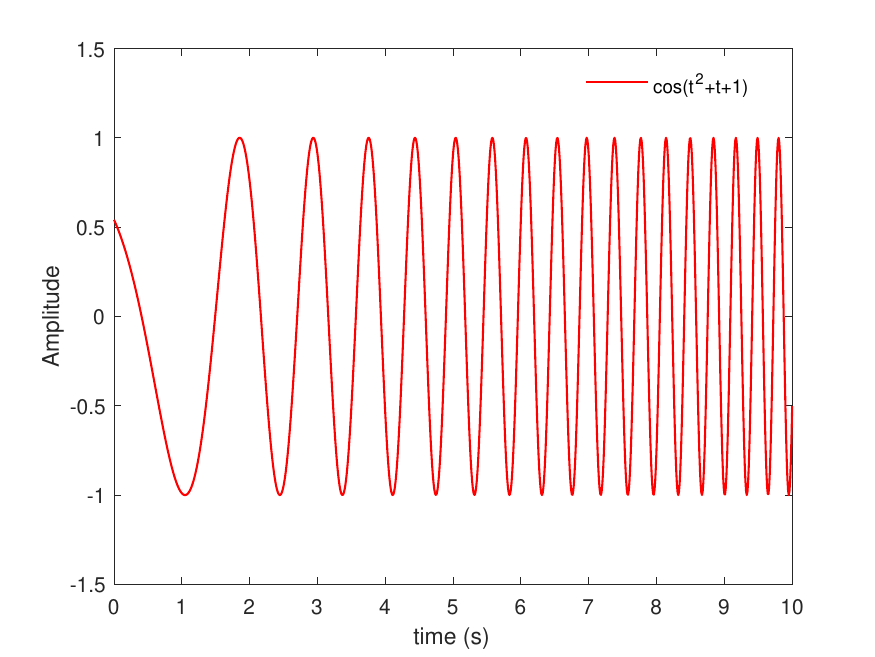}
	\caption{The plot of a cosine function with a quadratic trajectory in fractional dimensions.}
	\label{6}
\end{figure}


{\section{{Quantum Dynamics and the Geometry of Becoming:
Fractional Dimensional Time, Memory, and the Structure of Evolution}}}

Quantum mechanics describes physical change through unitary time evolution governed by a fixed first-order differential equation. While quantum states may exhibit nonlocal correlations, the structure of temporal evolution itself is assumed to be local, memoryless, and dimensionally fixed. In this study, we explore the conceptual consequences of relaxing this assumption. We examine the possibility that temporal evolution in quantum theory may be governed by a fractional or variable-dimensional structure, giving rise to what may be called a geometry of becoming. Within this perspective, memory is not introduced as a dynamical correction or environmental effect, but emerges as a structural property of time evolution itself where evolution itself is evolving.   \\

\section{Two-Channel Representation of the Schrödinger Equation}

In the geometric memory framework developed in this work, classical
kinematics can be represented through two channels: a local channel
described by ordinary derivatives and a geometric memory channel
described by fractional operators of variable order.
We now show that an analogous representation can be constructed
for quantum dynamics.

\subsection{Standard Schrödinger Dynamics}

The evolution of a closed quantum system is governed by the
time-dependent Schrödinger equation

\begin{equation}
i\hbar \frac{\partial}{\partial t} |\psi(t)\rangle
=
\hat H |\psi(t)\rangle ,
\end{equation}

where $\hat H$ is the Hamiltonian operator.
This equation represents a purely local and Markovian time evolution.

Within the geometric memory framework, however, temporal evolution
may also be represented through operators that incorporate
history-dependent memory effects.
This motivates the introduction of a two-channel representation.

\subsection{Two-Channel Quantum Representation}

Let $O_{t,N}^{\alpha(t)}$ denote a normalized fractional
operator of variable order $\alpha(t)$ acting on the temporal domain.

We introduce a decomposition of the quantum state into two channels,

\begin{equation}
|\psi(t)\rangle
=
C_1 |\psi_{\mathrm{loc}}(t)\rangle
+
C_2 |\psi_{\mathrm{mem}}(t)\rangle ,
\end{equation}

where

\[
C_1 , C_2 \ge 0, \qquad C_1 + C_2 = 1 .
\]

The two components represent distinct dynamical channels:

\begin{itemize}
\item Local channel
\end{itemize}

\begin{equation}
i\hbar
\frac{\partial}{\partial t}
|\psi_{\mathrm{loc}}(t)\rangle
=
\hat H
|\psi_{\mathrm{loc}}(t)\rangle .
\end{equation}

\begin{itemize}
\item Geometric memory channel
\end{itemize}

\begin{equation}
i\hbar
O_{t,N}^{\alpha(t)}
|\psi_{\mathrm{mem}}(t)\rangle
=
\hat H
|\psi_{\mathrm{mem}}(t)\rangle .
\end{equation}

The observable quantum state is therefore represented as a convex
combination of these two channels.

Combining both channels yields a unified evolution equation. Therefore, the unified two-Channel Schrödinger equation is:

\begin{equation}
i\hbar
\left[
C_1 \frac{\partial}{\partial t}
+
C_2 O_{t,N}^{\alpha(t)}
\right]
|\psi(t)\rangle
=
\hat H |\psi(t)\rangle .
\end{equation}

This equation defines the \emph{two-channel generator of quantum evolution}

\begin{equation}
\mathcal{D}_t^{(2)}
=
C_1 \partial_t
+
C_2 O_{t,N}^{\alpha(t)} .
\end{equation}

Thus the Schrödinger dynamics can be written compactly as

\begin{equation}
i\hbar
\mathcal{D}_t^{(2)}
|\psi(t)\rangle
=
\hat H |\psi(t)\rangle .
\end{equation}
To ensure that the two-channel representation reproduces the
standard Schrödinger dynamics, we impose the consistency condition

\begin{equation}
O_{t,N}^{\alpha(t)}
|\psi(t)\rangle
=
\frac{\partial}{\partial t}
|\psi(t)\rangle .
\end{equation}

Under this condition,

\begin{equation}
C_1\partial_t|\psi\rangle
+
C_2 O_{t,N}^{\alpha(t)}|\psi\rangle
=
(C_1+C_2)\partial_t|\psi\rangle
=
\partial_t|\psi\rangle .
\end{equation}

The unified equation therefore reduces exactly to the
standard Schrödinger equation.

Within this framework, quantum evolution admits two complementary
representations:

\begin{itemize}

\item a \textbf{local channel}, corresponding to instantaneous
Markovian dynamics;

\item a \textbf{geometric memory channel}, representing evolution
generated by  temporal memory encoded through the
variable-order operator $O_{t,N}^{\alpha(t)}$.

\end{itemize}

When $\alpha(t)=1$, the memory channel collapses to the standard
first-order derivative and the theory reduces to conventional
quantum mechanics.

For $\alpha(t)\neq1$, the same observable quantum evolution
can be represented through a nontrivial geometric-memory structure.
This representation parallels the two-channel description of
classical trajectories developed earlier in this work and
suggests a unified geometric interpretation of dynamics across
classical and quantum regimes.
\subsection{Path Integral Interpretation of the Geometric Memory Channel}

A natural physical motivation for the geometric memory channel arises
from the path integral formulation of quantum mechanics.
While the Schr\"odinger equation describes evolution through a local
first-order differential law in time, the Feynman formulation expresses
the transition amplitude as a sum over entire histories:

\begin{equation}
K(x_f,t_f;x_i,t_i)
=
\int \mathcal{D}[x]
\,\exp\!\left(\frac{i}{\hbar}S[x]\right),
\end{equation}

where \(S[x]\) is the classical action associated with a path \(x(t)\).
This representation is fundamentally nonlocal in time:
the amplitude between two events depends not on an instantaneous state
alone, but on an ensemble of trajectories connecting the initial and final
configurations.

Within the present framework, this distinction suggests a natural
two-channel interpretation of quantum evolution.

The path integral formulation indicates that quantum evolution may also
be viewed as arising from a temporally extended structure, since the
propagator incorporates contributions from all admissible histories.
This history dependence motivates the interpretation of the memory channel
as an effective geometric compression of path information into a
variable-order temporal operator:
\begin{equation}
\partial_t
\quad \rightsquigarrow \quad
O_{t,N}^{\alpha(t)} .
\end{equation}

In this interpretation, the operator \(O_{t,N}^{\alpha(t)}\) does not
represent an external environment or dissipative correction.
Rather, it encodes the influence of temporally extended history in a
reduced geometric form.
Thus, the memory channel is understood as a structural surrogate for
the nonlocality already implicit in the path integral.

\subsubsection{From sum over histories to variable-order memory.}
To make this idea more precise, consider the short-time propagation law
\begin{equation}
\psi(x,t+\Delta t)
=
\int K(x,t+\Delta t;x',t)\,\psi(x',t)\,dx' .
\end{equation}
In standard quantum mechanics, expansion in small \(\Delta t\) yields
the local Schr\"odinger equation.
However, if the effective temporal structure is not strictly local,
the propagation law may instead involve a weighted accumulation of prior
states:
\begin{equation}
\psi(t)
=
\psi(t_0)
+
\int_{t_0}^{t} \mathcal{K}(t,s)\,\hat H \psi(s)\,ds ,
\end{equation}
where \(\mathcal{K}(t,s)\) is a history kernel.
When this kernel has the scaling form
\begin{equation}
\mathcal{K}(t,s)
\sim
\frac{(t-s)^{\alpha(t)-1}}{\Gamma(\alpha(t))},
\end{equation}
the history dependence takes the form of a variable-order fractional
integral or derivative.
This yields an effective geometric-memory representation
\begin{equation}
i\hbar\, O_{t,N}^{\alpha(t)} \psi(t)
=
\hat H \psi(t).
\end{equation}

Therefore, the variable-order operator may be interpreted as a
coarse-grained encoding of path-history contributions.
The order trajectory \(\alpha(t)\) quantifies how strongly the present
state is distributed over a temporally extended region.
The limit \(\alpha(t)=1\) recovers the ordinary local derivative and
hence the standard Schr\"odinger equation.

In this view, the path integral and the geometric memory operator are not
independent descriptions but complementary ones.
The path integral is the full sum-over-histories description, while
\(O_{t,N}^{\alpha(t)}\) is an effective operator-level representation
of that temporal extension.
Thus, the geometric memory channel may be understood as the operator
shadow of the path integral's nonlocal structure.\\
\\
{\textbf{\Large {Part C: Mathematical Formulation}}}\\

In this section, we develop a mathematical framework to describe the fractional dimensional basis vectors and functions. Before delving into the specifics, we present some essential definitions related to fractional derivatives (FDr) and establish local FDr expressions for certain functions relevant to this study. It is important to note that our approach begins with an approximation of the connon component ($D^{\alpha}\Psi_0\approx \text{common term}$) to construct a physics-based framework from the perspective of fractional dimensions.

\section{\textbf{Definitions and General Framework}}

Here, we introduce a mathematical toolset applicable to this study, particularly focusing on the definitions of basis functions within fractional domains, while considering the realization of linear and fractional spaces.\\
\\
\noindent\textbf{Definition 1:} \textit{
Let $ \mathbb{R}^n$ denote an $ n $-dimensional Euclidean space. Then, a fractional space \( \mathbb{R}^{\alpha} \) can be defined as a continuous spectrum where \( \alpha \) is a real  number:}

\begin{equation}
\mathbb{R}^{\alpha} \quad \text{for } \alpha \in [n, n+1), \quad n \in \mathbb{Z}.
\end{equation}
Here, \( \alpha \) represents the fractional dimension that lies between the integer dimensions \( n \) and \( n+1 \).
A fractional space is a continuous spectrum that exists between and includes integer-dimensional spaces and can be generalized to complex .\\
\\
\noindent\textbf{Definition 2:} \textit{
Let \( \mathbf{e}_i \) represent the basis vectors in \( \mathbb{R}^n \). The basis functions in the fractional space \( \mathbb{R}^{\alpha} \) for \( \alpha \in (n, n+1) \) can be represented as a rotation \( \mathbf{R}_{\theta} \) of the integer-dimensional basis vectors \( \mathbf{e}_i \):}

\begin{equation}
\mathbf{e}_{\alpha} = \mathbf{R}_{\theta}(\mathbf{e}_i) \quad \text{for } \alpha \in (n, n+1),
\end{equation}
where \( \mathbf{R}_{\theta} \) is the rotation matrix parameterized by \( \theta \) that corresponds to the fractional dimension \( \alpha \).
In other words, if a fractional space is defined between two consecutive integer dimensions, then the basis functions within this space can be represented as rotations of the basis functions from the linear (integer) spaces.
\\

\noindent\textbf{Definition 3:} 
 \textit{The Gamma function is defined by
\[
\Gamma(z)=\int_{0}^{\infty} t^{z-1} e^{-t}\,dt, \qquad z>0,
\]
satisfying $\Gamma(z+1)=z\,\Gamma(z)$.  
It generalizes factorials, since $\Gamma(n+1)=n!$ for $n\in\mathbb{N}$, and plays a fundamental role in fractional calculus kernels and normalization factors \cite{Samko1993,Podlubny1999,Kilbas2006}.
}

\noindent\textbf{Definition 4:} \textit{
Let \( f(t) \) be a function defined on the interval \( (a, b) \). The fractional derivative of \( f(t) \) of order \( \alpha \in (0, +\infty) \) is denoted by \( D^{\alpha}f(t) \) and is defined as:}

\begin{equation}
D^{\alpha} f(t) = \frac{1}{\Gamma(1 - \alpha)} \int_a^t \frac{f(\tau)}{(t - \tau)^{\alpha}} \, d\tau, \quad \text{for } \alpha \in (0, 1),
\end{equation}
where \( \Gamma(\cdot) \) is the Gamma function, and the integral is understood in the sense of the Riemann-Liouville fractional derivative. For general \( \alpha \), this extends naturally to higher orders as:

\begin{equation}
{^{RL}D}^{\alpha} f(t) = \frac{1}{\Gamma(n - \alpha)} \left( \frac{d}{dt} \right)^n \int_a^t \frac{f(\tau)}{(t - \tau)^{\alpha - n + 1}} \, d\tau, \quad \text{for } \alpha \in (n-1, n), \quad n \in \mathbb{Z}^+.
\end{equation}\\
\begin{equation}
{^{C}D}^{\alpha} f(t) = \frac{1}{\Gamma(n - \alpha)}  \int_a^t \frac{1}{(t - \tau)^{\alpha + 1}}\frac{d^n f(\tau)}{dt^n} \, d\tau, \quad \text{for } \alpha \in (n-1, n), \quad n \in \mathbb{Z}^+.
\end{equation}\\
\noindent\textbf{Definition 5:} 
\textit{The fractional derivative consists of two components: a common term and an extra term, with the extra term differing based on the specific definition of the fractional derivative (FDr):}
\[
_{a}D_{t}^{\alpha}f(t) = \text{common term} + \text{extra term} = \text{local term} + \text{nonlocal term}.
\]
The extra term varies across different FDr definitions and is often expressed in terms of the generalized hypergeometric function ($F_{\alpha}(t))$\cite{herrmann2011fractional}. The common term, derived from the integer derivative, is referred to as the local fractional derivative, hence termed the local FDr.\\
We demonstrate and clarify the local and nonlocal part for some important functions as follows from \cite{RG} for both Riemann-
Liouville  and Caputo fractional derivatives:\\
\textbf{1) Exponential Function:}
Let $\alpha > 0$, $m = \lceil \alpha \rceil$, and $t_0 \in \mathbb{R}$. For any $\Omega \in \mathbb{C}$, $\left|\arg \Omega\right| < \pi$, and $t > t_0$, it holds \cite{RG}:

\begin{equation}
\begin{split}
^{RL} D^{\alpha}_{t_0} \Omega^t e^{\Omega (t - t_0)} = \Omega^\alpha e^{\Omega (t - t_0)} + F_\alpha(t - t_0; \Omega)\\
=local \ term \ (common \ term)+ nonlocal \ term \ (extra \ term)
\end{split}
\end{equation}
\begin{equation}
\begin{split}
^{C} D^{\alpha}_{t_0} \Omega^t e^{\Omega (t - t_0)} = \Omega^\alpha e^{\Omega (t - t_0)} + m F_{\alpha - m} (t - t_0; \Omega)
\end{split}
\end{equation}\\
where the local part (common part) based on both definitions  is $\Omega^\alpha e^{\Omega (t - t_0)}$.\\

\noindent \textbf{2) Trigonometric Function:}
Let \(\alpha > 0\), \(m = \lceil \alpha \rceil\), and \(\Omega \in \mathbb{R}\). Then for any \(t \geq t_0\), it holds for sinus and cosine function as follows \cite{RG}:

\begin{equation}
^{RL} D^{\alpha}_{t_0} \sin(\Omega (t - t_0)) = \Omega^\alpha \sin\left(\Omega (t - t_0) + \frac{\pi \alpha}{2}\right) + F_\alpha(t - t_0; i\Omega) - F_\alpha(t - t_0; -i\Omega)
\end{equation}

\begin{equation}
^{C} D^{\alpha}_{t_0} \sin(\Omega (t - t_0)) = \Omega^\alpha \sin\left(\Omega (t - t_0) + \frac{\pi \alpha}{2}\right) + i m \Omega^m F_{\alpha - m}(t - t_0; i\Omega) - (-1)^m F_{\alpha - m}(t - t_0; -i\Omega)
\end{equation}
where the local part (common part) based on both definitions  is $\Omega^\alpha \sin\left(\Omega (t - t_0) + \frac{\pi \alpha}{2}\right)$. For cosine funtion, we have:\\
\begin{equation}
^{RL} D^{\alpha}_{t_0} \cos(\Omega (t - t_0)) = \Omega^\alpha \cos\left(\Omega (t - t_0) + \frac{\pi \alpha}{2}\right) + \frac{F_\alpha(t - t_0; i\Omega) + F_\alpha(t - t_0; -i\Omega)}{2i}
\end{equation}

\begin{equation}
^{C} D^{\alpha}_{t_0} \cos(\Omega (t - t_0)) = \Omega^\alpha \cos\left(\Omega (t - t_0) + \frac{\pi \alpha}{2}\right) + \frac{i m \Omega^m F_{\alpha - m}(t - t_0; i\Omega) + (-1)^m F_{\alpha - m}(t - t_0; -i\Omega)}{2}
\end{equation}
where the local part (common part) based on both definitions  is $\Omega^\alpha \cos\left(\Omega (t - t_0) + \frac{\pi \alpha}{2}\right)$.\\

\noindent \textbf{3) Power Function:}
The basic results of a basic power function $t^n$ can be derived through direct methods
, Riemann–Liouville or Caputo, as following \cite{RG}:
\begin{equation}
^{RL} D^{\alpha}_{t_0} (t - t_0)^\beta =
\begin{cases}
\frac{\Gamma(\beta + 1)}{\Gamma(\beta - \alpha + 1)} (t - t_0)^{\beta - \alpha}, & \beta \in \{a - m, a - m + 1, \dots, a - 1\}, \\
0, & \text{otherwise}
\end{cases}
\end{equation}

\begin{equation}
^{C} D^{\alpha}_{t_0} (t - t_0)^\beta =
\begin{cases}
\frac{\Gamma(\beta + 1)}{\Gamma(\beta - \alpha + 1)} (t - t_0)^{\beta - \alpha}, & \beta > m - 1, \\
0, & \beta \in \{0, 1, \dots, m - 1\}, \\
\text{non-existing}, & \text{otherwise}
\end{cases}
\end{equation}

However in \cite{RG}, it is mentioned that this result is no longer valid for Caputo derivative once the m-th order derivative is evaluated the Caputo fractional derivative is no longer integrable.
Therefor a general form is provided as follows \cite{RG}:
Let \(\alpha > 0\) and \(m = \lceil \alpha \rceil\). Then for any \(k \in \mathbb{N}\)[]:

\begin{equation}
^{RL} D^{\alpha}_{t_0} t^k = 
\begin{cases} 
\sum_{l=0}^{k}\frac{k!}{(k - l)! \Gamma(l-\alpha+1)}t_0^{k - l}(t - t_0)^{l - \alpha}, & \text{if} \ k \geq \alpha, \\
0, & \text{otherwise}.
\end{cases}
\end{equation}

\begin{equation}
^{C}D^{\alpha}_{t_0} t^k =
\begin{cases}
0, & k < \alpha, \\
\sum_{l=m}^{k}\frac{k!}{(k - l)! \Gamma(l-\alpha+1)}t_0^{k - l}(t - t_0)^{k - \alpha}, &  \text{otherwise}.
\end{cases}
\end{equation}
where the common term is $\frac{\Gamma(k + 1)}{\Gamma(k - \alpha + 1)} (t - t_0)^{k - \alpha}$.\\

\noindent\textbf{Definition 6:} 
\textit{The normalized (local) fractional derivative (FDr) is obtained by dividing the FDr by a normalization factor deponed on function, which is a function of the order \( \alpha \). This normalization ensures that the basis functions maintain their normalized properties.}\\
For instance, let the normalization applies for local part. The Caputo fractional derivative for cosine function can expressed as follows: 
\begin{equation}
\begin{split}
   & ^{C} D^{\alpha}_{N,t_0} \cos(\Omega (t - t_0)) = \frac{1}{\Omega^\alpha } {^{C} D}^{\alpha}_{t_0} \cos(\Omega (t - t_0)) \\&=cos\left(\Omega (t - t_0) + \frac{\pi \alpha}{2}\right) +\frac{1}{\Omega^\alpha } \frac{i m \Omega^m F_{\alpha - m}(t - t_0; i\Omega) + (-1)^m F_{\alpha - m}(t - t_0; -i\Omega)}{2}
    \end{split}
\end{equation}
therefore, in general the normalized local FDr (NLFDr) for  cosine function $t=t_{0}$ is:
\begin{equation}
\begin{split}
   & D^{\alpha}_{N,L,t_0} \cos(\Omega (t - t_0)) =cos\left(\Omega (t - t_0) + \frac{\pi \alpha}{2}\right), 
    \end{split}
\end{equation}
and with the same approach for these functions, the NLFDr at $t=t_{0}$ are:
\begin{equation}
\begin{split}  
& D^{\alpha}_{N,L,t_0} sin(\Omega (t - t_0)) =cos(\Omega (t - t_0) + \frac{\pi \alpha}{2}),\\
&D^{\alpha}_{N,L,t_0} e^(\Omega (t - t_0)) =e^(\Omega (t - t_0) + \frac{\pi \alpha}{2}), \\
&D^{\alpha}_{N,L,t_0} t^{k}= (t - t_0)^{k - \alpha}.
\end{split} 
\end{equation}\\
\section*{Theorem 1: Commutativity and Symmetry of Normalized Local Fractional Derivative (NLFDr) Operators}

\subsection*{Statement of the Theorem}
Let \( D_{N,L}^{\alpha} \) denote the normalized local fractional derivative operator of order \( \alpha \). The following properties hold:

\begin{itemize}
    \item \textbf{Commutativity}: For two fractional orders \( \alpha_1 \) and \( \alpha_2 \), the normalized local fractional derivative operators commute, i.e.,
    \[
    D_{N,L}^{\alpha_1} D_{N,L}^{\alpha_2} = D_{N,L}^{\alpha_2} D_{N,L}^{\alpha_1} = D_{N,L}^{\alpha_1 + \alpha_2}.
    \]
    
    \item \textbf{Symmetry with Integral Operator}: Applying the normalized local fractional derivative operator followed by its corresponding integral operator of the same order returns the identity operation:
    \[
    I_{N,L}^{\alpha} D_{N,L}^{\alpha} = D_{N,L}^{\alpha} I_{N,L}^{\alpha} = \mathds{1},
    \]
 where \( I_{N,L}^{\alpha} \) is the normalized local fractional integral operator of order \( \alpha \).
\end{itemize}
\subsubsection*{\textbf{Proof of Commutativity}}

Assume \( f(t) \) is a function for which the fractional derivatives of orders \( \alpha_1 \) and \( \alpha_2 \) exist. By the definition of the normalized local fractional derivative operator, we have:

\[
D_{N,L}^{\alpha_1} (D_{N,L}^{\alpha_2} f(t)) = D_{N,L}^{\alpha_1 + \alpha_2} f(t).
\]

Similarly,

\[
D_{N,L}^{\alpha_2} (D_{N,L}^{\alpha_1} f(t)) = D_{N,L}^{\alpha_1 + \alpha_2} f(t).
\]

Thus, \( D_{N,L}^{\alpha_1} D_{N,L}^{\alpha_2} = D_{N,L}^{\alpha_2} D_{N,L}^{\alpha_1} = D_{N,L}^{\alpha_1 + \alpha_2} \).

\subsubsection*{\textbf{Proof of Symmetry with Integral Operator}}

Let \( f(t) \) be a function such that both \( D_{N,L}^{\alpha} \) and \( I_{N,L}^{\alpha} \) exist. The normalized local fractional integral operator is defined as the inverse operation to the normalized local fractional derivative. Therefore,

\[
I_{N,L}^{\alpha} (D_{N,L}^{\alpha} f(t)) = f(t).
\]

Similarly,

\[
D_{N,L}^{\alpha} (I_{N,L}^{\alpha} f(t)) = f(t).
\]

Hence,\( I_{N,L}^{\alpha} D_{N,L}^{\alpha} = D_{N,L}^{\alpha} I_{N,L}^{\alpha} = \mathds{1} \).

\subsection{\textbf{Example}}
Let the normalized local FDr operator be applied to cosine function. This operator exhibits commutative and symmetry properties for these functions, as demonstrated below:
\begin{equation}
D_{\rm N,L}^{\alpha_1 }D_{\rm N,L}^{\alpha_2 } = D_{\rm N,L}^{\alpha_1 + \alpha_2 },
\label{49}
\end{equation}
\begin{equation}
D_{\rm N,L}^{\alpha }I_{\rm N,L}^{\alpha } = \mathds{1},
\label{50}
\end{equation}
Starting with the cosine function on the left-hand side of Eq.~\eqref{49}:

\begin{equation}
D_{\rm N,L}^{\alpha_1 }D_{\rm N,L}^{\alpha_2 }\left\{\cos(\omega S)\right\} = D_{\rm N,L}^{\alpha_1 }
\left\{\cos\left(\omega S + \frac{\pi}{2}\alpha_2\right)\right\} = 
\cos\left(\omega S + \frac{\pi}{2}\alpha_2 + \frac{\pi}{2}\alpha_1\right),
\end{equation}

For the right-hand side:

\begin{equation}
D_{\rm N,L}^{\alpha_1 + \alpha_2 }\left\{\cos(\omega S)\right\} = \cos\left(\omega S + \frac{\pi}{2}(\alpha_1 + \alpha_2)\right).
\end{equation}

Thus, the two sides are equivalent. This reasoning also applies to the sine, complex basis, and power series functions.

To prove Eq.~\eqref{50}, we first note that the fractional integral of order $\alpha$ ($I_{\rm N,L}^{\alpha }$) for any function $f$ can be expressed based on the FDr:

\begin{equation}
I_{\rm N,L}^{\alpha}f = D_{\rm N,L}^{-\alpha}f.
\end{equation}

Then, we have:

\begin{equation}
D_{\rm N,L}^{\alpha }I_{\rm N,L}^{\alpha }\left\{\cos(\omega S)\right\} = D_{\rm N,L}^{\alpha }
\left\{\cos\left(\omega S - \frac{\pi}{2}\alpha\right)\right\} = 
\cos\left(\omega S - \frac{\pi}{2}\alpha + \frac{\pi}{2}\alpha\right) = \cos(\omega S),
\end{equation}

This property holds for other proposed functions as well.

\subsection{Fractional Dimensional Basis Vector}

An alternative approach to achieve the results presented in Eq.~\eqref{DS} involves considering the relationship between the coordinates $S$ and $S_{\alpha}$, expressed as a superposition of the FDr (order $\alpha$) of $S$ on one hand, and the fractional integral (order $\beta = 1 - \alpha$) of 1 on the other. It is important to note that the sum of the fractional orders must equal one to correspond to a 90-degree rotation:

\begin{equation}
\begin{split}
S_{\alpha} = (1-\alpha)D_{\rm N,L}^{\alpha}\{S\} + \alpha I_{\rm N,L}^{\beta }\{1\} = (1-\alpha)D_{\rm N,L}^{\alpha}\{S\} + \alpha D_{\rm N,L}^{-\beta }\{1\}.
\label{4}
\end{split}
\end{equation}

The coefficients of $D_{\rm N,L}^{\alpha}$ and $I_{\rm N,L}^{\alpha}$ act as weight factors, $(1-\alpha)$ and $(\alpha)$, respectively. These weights ensure that when $\alpha = 0$, the coordinates $S$ and $S_{\alpha}$ coincide, and when $\alpha = 1$, a 90-degree rotation in fractional dimensional space occurs, resulting in $S_{\alpha}$ as:

\begin{equation}
\begin{split}
S_{\alpha} = (1-\alpha)S^{1-\alpha} + \alpha S^{1-\alpha} = S^{1-\alpha}.
\end{split}
\end{equation}

\subsection{Example}

To illustrate, the 3D plot of the FDr of the function $f(x) = x$ over the interval $x \in [0,10]$ and $\alpha \in [0,1]$ is depicted in Fig.~\ref{fig7}.

\begin{figure}[H]
    \centering
    \includegraphics[width=.8\linewidth]{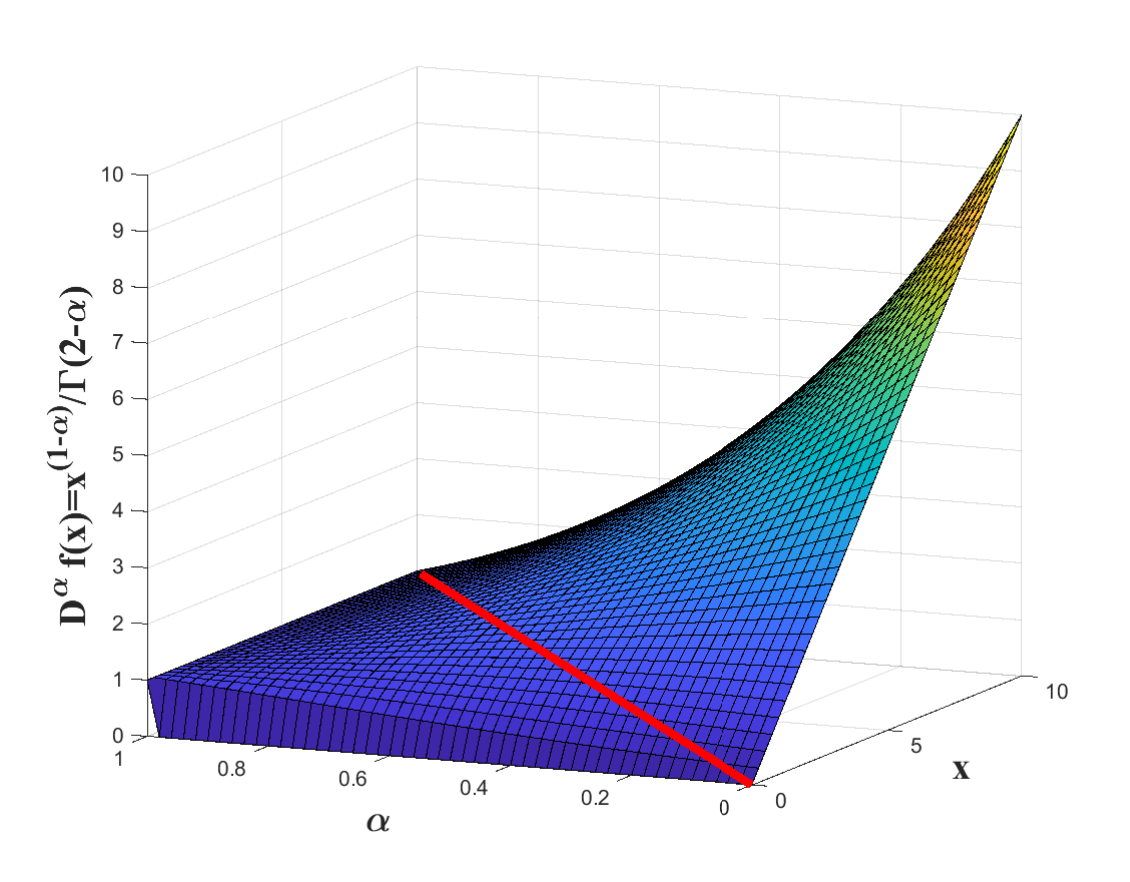}
    \caption{FDr of the function $f(x) = x$ over the interval $[0,10]$ and $\alpha$ between 0 and 1.}
    \label{fig7}
\end{figure}
\noindent Next, consider following a trajectory along the red line in Fig.~\ref{fig7}. The red line corresponds to the trajectory $\alpha = 0.1x$, which leads to the FDr plot along this trajectory, as shown in Fig.~\ref{fig8}. This implies that for each point $x$, there exists a specific $\alpha$ that forms the function $\alpha(x)$.

\begin{figure}[h!]
    \centering
    \includegraphics[width=.8\linewidth]{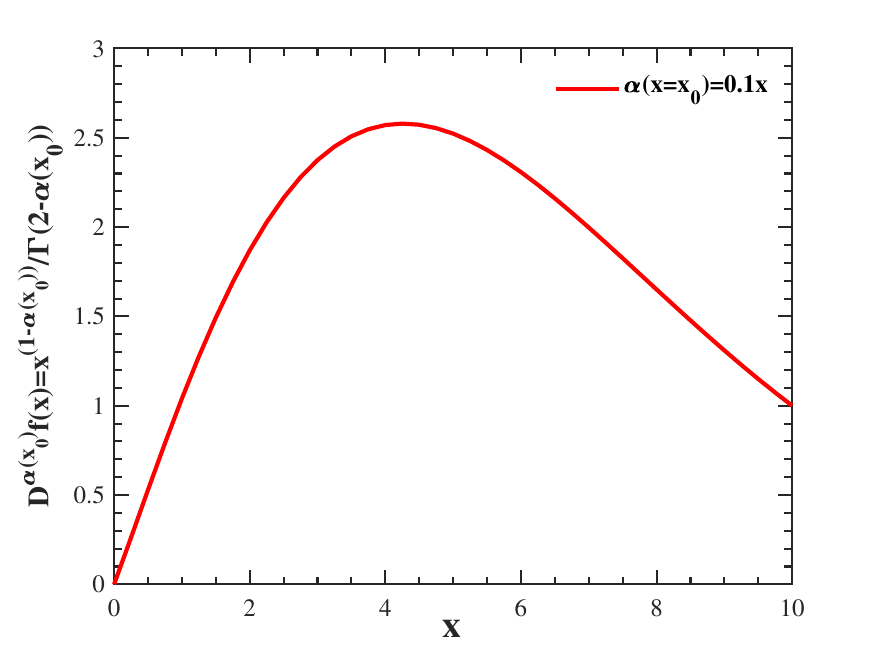}
    
    \caption{FDr of the function $f(x) = x$ over the interval $[0,10]$ for the trajectory $\alpha = 0.1x$. Note that for each point $x$, there is a corresponding constant value of $\alpha$.}
    \label{fig8}
\end{figure}


\section{
Representation of Functions by Fractional-Dimensional Tangent Lines
through Fractional Operators of Variable Order}

\subsection{Introduction}
\label{sec:intro}
The mathematical representation of functions has long been central in analysis and applications. 
Classical schemes—Taylor series, Fourier expansions, and power series 
\cite{HMF,FS,TS-PS,GDRS}—use integer-order derivatives and are effective for smooth or analytic functions. 
While powerful, such approaches are inherently local: they rely on finitely many integer-order derivatives at a point and cannot capture long-range memory or nonlocal effects.

Fractional calculus, which extends differentiation and integration to arbitrary real (or complex) orders, 
has been developed systematically in monographs \cite{Samko1993,Kilbas2006,Diethelm2010,Mainardi2010,Herrmann2018} 
and reviewed in \cite{RG}. 
Fractional operators combine local slopes with nonlocal memory terms and are widely applied in viscoelasticity, anomalous diffusion, signal processing, and dynamical systems \cite{West2017,Tarasov2010,Povstenko2015,Mainardi2010}.  

Within this framework, the classical notion of a tangent line can be extended to 
\emph{fractional dimensions}, leading to the concept of \emph{fractional tangent lines}. 
These objects combine local and nonlocal characteristics of fractional operators 
and provide a richer geometric interpretation of function behavior 
\cite{DA-TLFDS,DS-Roll,DS-opt,DS-FFS}.

In this article we establish a representation of functions via fractional tangent lines using general fractional operators. 
We introduce the necessary preliminaries, 
prove existence (Lemma~\ref{lem:line}), describe two representation strategies (Theorems~\ref{thm:sc1}--\ref{thm:sc2}), 
with an geometrical explanation.

\subsection{Preliminaries}
We proceed by recalling the main notions and operators introduced in Definitions 1–6. Let $\alpha>0$ and let $I^\alpha$ denote the fractional integral of order $\alpha$,
\[
(I^{\alpha} f)(t)=\frac{1}{\Gamma(\alpha)}\int_a^t (t-\tau)^{\alpha-1} f(\tau)\,d\tau.
\]
We use the collective notation
\[
\mathcal{O}^\alpha \in \bigl\{\,^{RL}D^\alpha,\,^{C}D^\alpha,\,I^\alpha\bigr\}.
\]
Standing assumptions (SA) can be expressed as follows:
\begin{framed}
\noindent\textbf{(SA-A) Integral-based representation.}
Fix $a\in\mathbb{R}$ and $\tau\in(a,b)$. Assume $f\in L^1_{\text{loc}}(a,b)$.
We define the fractional slope by $(\,^{RL}\!I^\alpha f)(\tau)$ with $\alpha>0$.
If, in addition, $f\ge 0$ a.e.\ on $(a,\tau)$ and $f\not\equiv 0$, then
$\alpha\mapsto(\,^{RL}\!I^\alpha f)(\tau)$ is continuous and strictly increasing on $(0,\infty)$.

\smallskip
\noindent\textbf{(SA-B) Derivative-based representation.}  
Fix $a\in\mathbb{R}$ and $\tau\in(a,b)$.  
Assume f is absolutely continuous (AC), $f\in AC^n([a,\tau])$, so that $f^{(n)}\in L^1(a,\tau)$.  
For $n-1<\alpha<n$, define the fractional slope by $(\,^{C}\!D^\alpha f)(\tau)$ (or $(\,^{RL}\!D^\alpha f)(\tau)$).  
If $f^{(n)}\ge 0$ on $[a,\tau]$ and $f^{(n)}\not\equiv 0$, then  
\[
\alpha \;\mapsto\; (\,^{C}\!D^\alpha f)(\tau)
\]
is continuous and strictly decreasing on $(n-1,n)$.
\end{framed}

\begin{definition}[Smoothness requirement]\label{def:smooth}
Let $f:(a,b)\to R$ be differentiable up to order $n+1$.  
This assumption ensures that the fractional operators in Definition~\ref{def:operators} are well defined and the underlying integrals converge in standard settings \cite{Kilbas2006,Diethelm2010}.
\end{definition}

\begin{definition}[Chord slope]
For two points $(t_0,f(t_0))$ and $(t,f(t))$ with $t\neq t_0$,
\[
m(t)=\frac{f(t)-f(t_0)}{t-t_0}.
\]
This is the target slope that we realize via fractional operators.
\end{definition}

\begin{definition}[Fractional tangent line]
Let $f$ satisfy Definition~\ref{def:smooth} and fix $(t_0,f(t_0))$.  
A \emph{fractional tangent line} to $f$ is
\[
L_\alpha(t)=(\mathcal{O}^\alpha f)(\tau)\,(t-t_0)+f(t_0),
\]
where $\mathcal{O}^\alpha$ is a fractional derivative or integral, and $\tau$ is either $t_0$ (fixed anchor) or $t$ (moving anchor). 
For background on the behavior of these operators.
\end{definition}

\begin{proposition}[Continuity and monotonicity of $\alpha\mapsto(\mathcal{O}^\alpha f)(\tau)$]\label{prop:phi-properties}
Assume \textup{(SA)} and fix $\tau\in(a,b)$.
\begin{enumerate}[label=(\roman*)]
\item \textbf{RL integral.} If $f\in L^1(a,\tau)$ and $f\ge0$ a.e.\ on $(a,\tau)$, then the map
\[
\alpha\mapsto (\,^{RL}\!I^{\alpha}f)(\tau)
=\frac{1}{\Gamma(\alpha)}\!\int_{a}^{\tau} (\tau-\xi)^{\alpha-1}f(\xi)\,d\xi
\]
is continuous and \emph{strictly increasing} on $(0,\infty)$ provided $f\not\equiv0$ on a set of positive measure.

\item \textbf{Caputo derivative of order $\alpha\in(0,1)$.} 
If $f\in C^1([a,\tau])$ and $f'\ge0$ on $[a,\tau]$, then
\[
\alpha\mapsto (\,^{C}\!D^{\alpha}f)(\tau)
=\frac{1}{\Gamma(1-\alpha)}\!\int_{a}^{\tau} (\tau-\xi)^{-\alpha} f'(\xi)\,d\xi
\]
is continuous and \emph{strictly decreasing} on $(0,1)$, provided $f'\not\equiv0$ on a set of positive measure.

\item \textbf{RL derivative of order $\alpha\in(0,1)$.}
If $f\in AC([a,\tau])$ and $f\ge0$ is nondecreasing on $[a,\tau]$, then
\[
\alpha\mapsto (\,^{RL}\!D^{\alpha}f)(\tau)
=\frac{1}{\Gamma(1-\alpha)}\frac{d}{d\tau}\int_{a}^{\tau} (\tau-\xi)^{-\alpha} f(\xi)\,d\xi
\]
is continuous on $(0,1)$ and \emph{nonincreasing}. It is strictly decreasing if $f$ is not a.e.\ constant near $\tau$.
\end{enumerate}
\end{proposition}

\begin{proof}
(i) Continuity follows from dominated convergence using the kernel 
$k_\alpha(s)=\frac{s^{\alpha-1}}{\Gamma(\alpha)}$ which depends continuously on $\alpha>0$ and is dominated on $s\in(0,\tau-a)$ by $s^{\alpha_0-1}$ for any fixed $\alpha_0>0$. 
Monotonicity: for $f\ge0$, the kernel $s^{\alpha-1}$ increases pointwise in $\alpha$, whence the integral increases; division by $\Gamma(\alpha)$ preserves strict increase because $\Gamma$ is log-convex and positive on $(0,\infty)$ while the $s^{\alpha-1}$–effect dominates (a standard argument via differentiation under the integral sign shows $\partial_\alpha\big[(^{RL}\!I^\alpha f)(\tau)\big]>0$ unless $f\equiv0$ a.e.).

(ii) Write 
$(^{C}\!D^\alpha f)(\tau)=\int_{0}^{\tau-a} \frac{u^{-\alpha}}{\Gamma(1-\alpha)}\, f'(\tau-u)\,du$ 
with $u=\tau-\xi$.
For $f'\ge0$, the kernel $u^{-\alpha}/\Gamma(1-\alpha)$ \emph{decreases} in $\alpha\in(0,1)$; dominated convergence gives continuity and strict decrease when $f'\not\equiv0$.

(iii) Use the representation 
$(^{RL}\!D^\alpha f)(\tau)=\frac{1}{\Gamma(1-\alpha)}\!\int_{0}^{\tau-a} u^{-\alpha} f'(\tau-u)\,du$
valid for $f\in AC$; this reduces to (ii). Nonincrease and strict decrease follow as above.
\end{proof}

\begin{proposition}[Classical tangent as a limit]\label{prop:classical-limit}
Assume \textup{(SA)} with $f\in C^1$ near $t_0$ and fix $\tau=t_0$.
For $\,^{C}\!D^\alpha$ with $\alpha\in(0,1)$,
\[
\lim_{\alpha\to 1} (\,^{C}\!D^\alpha f)(t_0) = f'(t_0).
\]
Consequently,
\[
\lim_{\alpha\to 1}\Big(f(t_0)+(\,^{C}\!D^\alpha f)(t_0)(t-t_0)\Big)
= f(t_0)+f'(t_0)(t-t_0),
\]
so the fractional tangent converges to the classical tangent line as $\alpha\to 1^{-}$.
\end{proposition}

\begin{proof}
Use  
$(^{C}\!D^\alpha f)(t_0)=\frac{1}{\Gamma(1-\alpha)}
\int_{0}^{t_0-a} u^{-\alpha}\, f'(t_0-u)\,du$ , and the well-known identity \\
$\lim_{\varepsilon\to 0}\frac{\varepsilon}{\Gamma(\varepsilon)}\int_{0}^{\delta} u^{\varepsilon-1}\,g(u)\,du=g(0)$ for continuous $g$; set $\varepsilon=1-\alpha$ and $g(u)=f'(t_0-u)$.
\end{proof}

\begin{lemma}[Linearity and affine invariance]\label{lem:affine}
For any $\alpha>0$ and scalars $c_1,c_2$,
\[
\mathcal{O}^\alpha[c_1 f + c_2 g]=c_1\,\mathcal{O}^\alpha f + c_2\,\mathcal{O}^\alpha g.
\]
In particular, for $f(t)=A(t-t_0)+B$,
\[
(\mathcal{O}^\alpha f)(\tau)=A\,(\mathcal{O}^\alpha (t-t_0))(\tau)+B\,(\mathcal{O}^\alpha 1)(\tau),
\]
so the slope condition $(\mathcal{O}^{\alpha(t)} f)(\tau)=A$ reduces to two precomputable kernels $(\mathcal{O}^\alpha (t-t_0))(\tau)$ and $(\mathcal{O}^\alpha 1)(\tau)$.
\end{lemma}

\begin{proof}
Linearity is standard for RL/Caputo derivatives and RL integrals. The identity follows by expansion.
\end{proof}

\subsubsection{Existence of a fractional tangent line}

\begin{lemma}[Existence of a fractional tangent line]\label{lem:line}
Let $f:(a,b)\to R$ satisfy Definition~\ref{def:smooth}.  
Fix $(t_0,f(t_0))$ and let $(t_n,f(t_n))$ be another point with $t_n\neq t_0$.  
Then there exists $\alpha_i>0$ and $\mathcal{O}^{\alpha_i}\in\{\,^{RL}D^{\alpha_i},\,^{C}D^{\alpha_i},\,^{RL}I^{\alpha_i}\}$, evaluated at $t_i\in\{t_0,t_n\}$, such that
\begin{equation}\label{eq:lemma}
f(t_n)=(\mathcal{O}^{\alpha_i} f)(t_i)\,(t_n-t_0)+f(t_0).
\end{equation}
\end{lemma}

\begin{proof}
The chord slope is $m=\frac{f(t_n)-f(t_0)}{t_n-t_0}$, hence $f(t_n)=m\,(t_n-t_0)+f(t_0)$.  
By continuity (and often monotonicity) of $\alpha\mapsto (\mathcal{O}^\alpha f)(t_i)$ (cf.\ \cite{Samko1993,Podlubny1999,Kilbas2006,Diethelm2010,Garrappa2019}), 
there exists $\alpha_i>0$ with $(\mathcal{O}^{\alpha_i} f)(t_i)=m$ (for $m$ in the operator’s range).  
Substitution yields \eqref{eq:lemma}.
\end{proof}

\begin{remark}
Lemma~\ref{lem:line} shows the chord slope can be realized by a suitable fractional operator at either endpoint, producing a \emph{fractional tangent line}.  
If $\alpha=1$ and $\mathcal{O}^\alpha=D^1$, one recovers the classical tangent
\[
f(t)=f'(t_0)(t-t_0)+f(t_0).
\]
\end{remark}
Figure~\ref{TL1} illustrates Lemma~\ref{lem:line}.  
The same chord slope $m$ can be matched either by $(\mathcal{O}^{\alpha_0}f)(t_0)$ or by $(\mathcal{O}^{\alpha_n}f)(t_n)$.  
Thus fractional tangent lines at $t_0$ and $t_n$ coincide.

\begin{figure}[h!]
\centering
\includegraphics[width=1\linewidth]{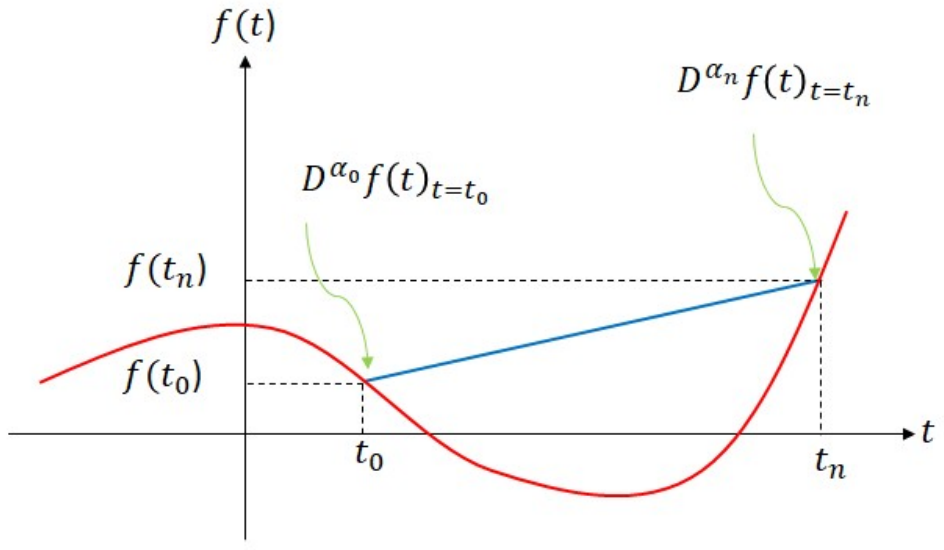}
\caption{ A fractional-dimensional tangent line at $(t_0,f(t_0))$ with order $\alpha_0$ and at $(t_n,f(t_n))$ with order $\alpha_n$, both reproducing the same chord slope $m$. }
\label{TL1}
\end{figure}

\subsubsection{Existence and uniqueness of the order map \texorpdfstring{$\alpha(t)$}{alpha(t)}}

\begin{corollary}[Practical ranges]\label{cor:practical}
Under the hypotheses of Proposition~\ref{prop:phi-properties}:
\begin{itemize}
\item For $^{RL}I^\alpha$ with $f\ge0$, one may take $I=(0,\infty)$.
\item For $\,^{C}\!D^\alpha$ (or $\,^{RL}\!D^\alpha$ via AC–reduction) with $f'\ge0$, one may take $I=(0,1)$, with $\Phi$ strictly decreasing.
\end{itemize}
Therefore, bracketing for the bisection/Newton scheme is always possible once $m(t)$ is observed to lie in the corresponding range.
\end{corollary}
\begin{theorem}[Well-posedness of the order map $\alpha(t)$]\label{thm:alpha-wellposed}
Assume \textup{(SA)}, fix $\tau\in\{t_0,t\}$, and define 
$\Phi_\tau^{(\mathcal{O})}(\alpha):=(\mathcal{O}^\alpha f)(\tau)$.
Let $m(t)=\frac{f(t)-f(t_0)}{t-t_0}$ for $t\neq t_0$.
Suppose there is an interval $I\subset R$ such that 
\begin{enumerate}[label=(\alph*)]
\item $\Phi_\tau^{(\mathcal{O})}$ is continuous and strictly monotone on $I$;
\item $m(t)\in \Phi_\tau^{(\mathcal{O})}(I)$.
\end{enumerate}
Then the scalar equation $\Phi_\tau^{(\mathcal{O})}(\alpha(t))=m(t)$ has a unique solution $\alpha(t)\in I$.
If, in addition, $\Phi_\tau^{(\mathcal{O})}\in C^1(I)$ with $\inf_{\alpha\in I}|(\Phi_\tau^{(\mathcal{O})})'(\alpha)|=\kappa>0$, then the inverse $m\mapsto \alpha(m)$ is Lipschitz with constant $\le 1/\kappa$.
\end{theorem}

\begin{proof}
Continuity and the range condition  yield existence by the intermediate value theorem.
Strict monotonicity  implies injectivity, hence uniqueness.
If \(\Phi_{\tau}^{(\mathcal{O})}\in C^1\) with derivative bounded away from zero, then the inverse function theorem gives differentiability of the inverse with \(|(\alpha(m))'|\le 1/\kappa\), hence the Lipschitz bound.
\end{proof}

\begin{remark}\label{rem:mono}
For many simple classes (e.g.\ nonnegative polynomials near $\tau>0$ with lower terminal $a=0$):
\begin{itemize}[leftmargin=1.2em,topsep=2pt,itemsep=1pt]
\item \(\alpha\mapsto (\,^{C}\!D^{\alpha}f)(\tau)\) is continuous and strictly \emph{decreasing} on \((0,1)\);
\item \(\alpha\mapsto (\,^{RL}\!I^{\alpha}f)(\tau)\) is continuous and strictly \emph{increasing} on \((0,\infty)\).
\end{itemize}
Thus one may choose $I=(0,1)$ for derivatives and $I=(0,\infty)$ for integrals, bracket \(m(t)\), and apply Theorem~\ref{thm:alpha-wellposed}. 
\end{remark}

\begin{lemma}[Realizing the chord slope by a fractional operator]\label{lem:realize-slope}
Assume \textup{(SA)} and fix $t_0\in(a,b)$.
For any $t\in(a,b)\setminus\{t_0\}$ let $m(t)=\frac{f(t)-f(t_0)}{t-t_0}$.
Choose $\mathcal{O}^\alpha\in\{\,^{RL}\!D^\alpha,\,^{\
C}\!D^\alpha,\,^{RL}\!I^\alpha\}$ and $\tau\in\{t_0,t\}$ so that the hypotheses of Proposition~\ref{prop:phi-properties} hold and let $I$ be the corresponding monotonicity interval. 
If $m(t)\in \Phi_\tau^{(\mathcal{O})}(I)$, there exists a \emph{unique} $\alpha(t)\in I$ such that
\[
f(t)=f(t_0)+(\mathcal{O}^{\alpha(t)}f)(\tau)\,(t-t_0).
\]
\end{lemma}

\begin{proof}
By Theorem~\ref{thm:alpha-wellposed} there is a unique $\alpha(t)\in I$ with $(\mathcal{O}^{\alpha(t)}f)(\tau)=m(t)$. Substitute in the right-hand side.
\end{proof}

\subsubsection{Two representation strategies}
\label{sec:two-scenarios}
Figure~\ref{TL2} shows the two scenarios: left = fixed anchor at $t_0$, right = moving anchor at $t$.

\begin{figure}[h!]
\centering
\includegraphics[width=1\linewidth]{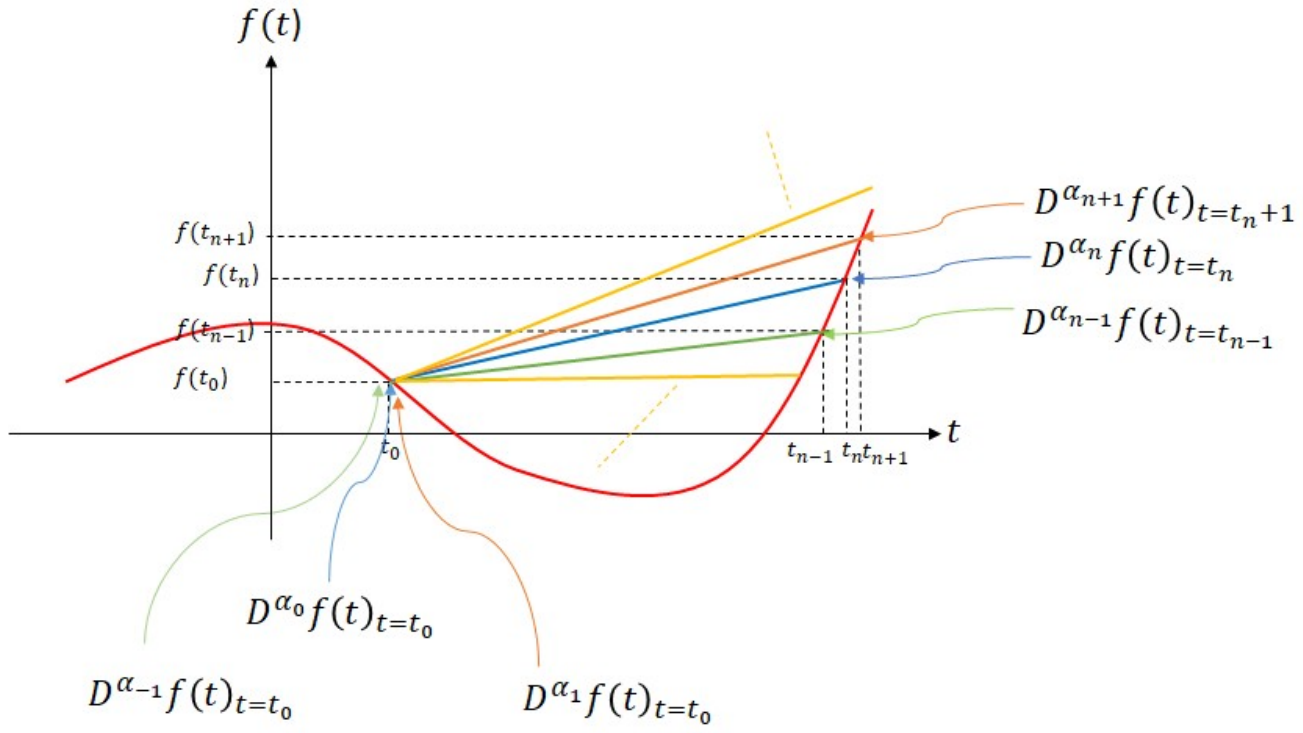}
\caption{ \emph{Left:} fixed anchor $(t_0,f(t_0))$; choose $\alpha(t)$ so that $(\mathcal{O}^{\alpha(t)}f)(t_0)=m(t)$, then $f(t)=(\mathcal{O}^{\alpha(t)}f)(t_0)(t-t_0)+f(t_0)$. \emph{Right:} moving anchor at $t$; choose $\alpha(t)$ so that $(\mathcal{O}^{\alpha(t)}f)(t)=m(t)$, then $f(t)=(\mathcal{O}^{\alpha(t)}f)(t)(t-t_0)+f(t_0)$.}
\label{TL2}
\end{figure}

\begin{itemize}
\item \textbf{Scenario 1 (fixed anchor).} Fix $(t_0,f(t_0))$ and connect all $(t,f(t))$ by choosing $\alpha(t)$ so that $(\mathcal{O}^{\alpha(t)}f)(t_0)=m(t)$. This defines a trajectory in fractional space anchored at $t_0$ (left panel of Figure~\ref{TL2}).
\item \textbf{Scenario 2 (moving anchor).} For each $(t,f(t))$, choose $\alpha(t)$ so that $(\mathcal{O}^{\alpha(t)}f)(t)=m(t)$, then connect back to $(t_0,f(t_0))$ (right panel of Figure~\ref{TL2}).
\end{itemize}

\begin{theorem}[Scenario 1: fixed anchor]\label{thm:sc1}
Let $f$ satisfy Definition~\ref{def:smooth}.  
Then
\begin{equation}\label{eq:sc1}
    f(t) = (\mathcal{O}^{\alpha(t)} f)(t_0)\,(t-t_0) + f(t_0),
\end{equation}
where $\alpha(t)>0$ is chosen so that $(\mathcal{O}^{\alpha(t)} f)(t_0)=m(t)=\tfrac{f(t)-f(t_0)}{t-t_0}$.
\end{theorem}

\begin{proof}
Apply Lemma~\ref{lem:line} with evaluation at $t_0$; substitute $(\mathcal{O}^{\alpha(t)} f)(t_0)=m(t)$.
\end{proof}

\begin{theorem}[Scenario 2: moving anchor]\label{thm:sc2}
Under the same assumptions,
\begin{equation}\label{eq:sc2}
    f(t) = (\mathcal{O}^{\alpha(t)} f)(t)\,(t-t_0) + f(t_0),
\end{equation}
where $\alpha(t)>0$ is chosen so that $(\mathcal{O}^{\alpha(t)} f)(t)=m(t)$.
\end{theorem}

\begin{proof}
Apply Lemma~\ref{lem:line} with evaluation at $t$; substitute $(\mathcal{O}^{\alpha(t)} f)(t)=m(t)$.
\end{proof}

\section{\textbf{Dimensional Transformation }}
Transformation of function from one space to another space is a promising method to reach an easier
analysis and more information in different view \cite{AA,AG, BY,MR, OF}. In principle, it is a mapping from real world to other
virtual world. For instance, Fourier transformation is a mapping from this real world to inverse world, time
to frequency or position to momentum. However, if considered Fourier transformation brings a function
from time domain to frequency domain, time and frequency as a dimension, however; intrinsically
dimension is inside time (space) or frequency (momentum). In other words, when we talk about time, automatically it is
considered as a dimension. In this point of view, brings out this concept that time is one domain and
dimension also another domain which has been coupled together. Thus, it is mentioned that dimension
itself is a dimension. In other words, when analyzed the function of f(t) means f (t, $\alpha$= 0)
in an original
dimension of function.
In this regard, dimensional transformation is defined as shown in the following the ma
pping of function in fractional dimension space.

\subsection{Dimensional Transformation Definitions}

Dimensional transformation can be introduced in two forms, each of which extends the analysis of a function into fractional dimensional space.

\subsubsection{First Type of Dimensional Transformation}

The first type of dimensional transformation, denoted as $DT_1$, is defined as:

\begin{equation}
DT_1\{f(t)\} := F(\alpha) = \int_{-\infty}^{\infty} \mathcal{O}^\alpha f(t) \, dt,
\end{equation}
where $\mathcal{O}^\alpha$ represents a fractional operator of order $\alpha$, and $\alpha \in \mathbb{R}(\mathbb{C})$ is the fractional dimension and can be a variable order. This operator is analogous to fractional derivatives and integrals, extending the domain of $f(t)$ into fractional dimensions.

\subsubsection{Second Type of Dimensional Transformation}

The second type of dimensional transformation, denoted as $DT_2$, is given by:

\begin{equation}
DT_2\{f(t)\} := F(\alpha) = \int_{-\infty}^{\infty} f(t) \, t^{-\alpha} \, dt.
\end{equation}

In this case, the function is transformed using the kernel $t^{-\alpha}$, which allows us to explore the properties of $f(t)$ in a different dimensional framework. The value of $\alpha$ can be any real number, extending the function into non-integer dimensions.

\begin{figure}[h]
  \includegraphics[width=1\linewidth]{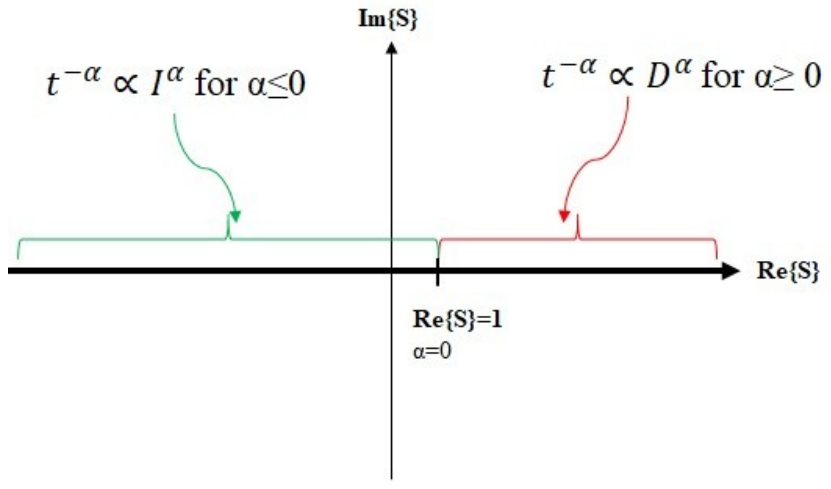}  
  \caption{Complex plane s with position of $\alpha$  on real s for dimensional transformation}
  \label{fig:two step process }
\end{figure}
There are two approaches to connect the $\mathcal {DT}$ to Mellin's transformation as follows:\\
Approach I:
\begin{equation}
\begin{split}
&F(\alpha)=\int_{-\infty}^{+\infty} f(t) t^{-\rm\alpha} dt =\int_{-\infty}^{0} f(t) t^{-\rm\alpha} dt+\int_{0}^{+\infty} f(t) t^{-\rm\alpha} dt\\&=- \int_{0 }^{-\infty} f(t) t^{-\rm\alpha} dt+\int_{0 }^{+\infty } f(t) (t)^{-\rm\alpha}  dt=\int_{0 }^{+\infty } f(-t) (-t)^{-\rm\alpha}  dt\\&+\int_{0 }^{+\infty } f(t) (t)^{-\rm\alpha}  dt=(-1)^{\rm\ s-1}\int_{0 }^{+\infty } f(-t) (t)^{\rm\ s-1}  dt+\int_{0 }^{+\infty } f(t) (t)^{\rm s-1} dt\\&=(-1)^{\rm\ s-1}\mathcal M\big\{ f(-t)\big\}+ \mathcal M\big\{ f(t)\big\}
\end{split}
\end{equation}
where $\mathcal M$ is the Mellin transformation\cite{ OF}.\\
Approach II:
\begin{equation}
\begin{split}
F(\alpha)=\int_{-\infty}^{+\infty} f(t) t^{-\rm\alpha} dt =\int_{0 }^{+\infty } f(-t) (-t)^{-\rm\alpha}  dt+\int_{0 }^{+\infty } f(t) t^{-\rm\alpha}  dt\\
\end{split}
\end{equation}

by applying $t=\frac{1}{T}$ and  $dt=\frac{-dT}{T^2}$ 

\begin{equation}
\begin{split}
\int_{+\infty}^{0 } f(-\frac{1}{T}) (-\frac{1}{T})^{-\rm\alpha}     \frac{-dT}{T^2}+\int_{+\infty }^{0 } f(\frac{1}{T}) (\frac{1}{T})^{-\rm\alpha} \frac{-dT}{T^2}\\
=(-1)^{-\rm\alpha}\int_{0 }^{+\infty } \frac{f(-\frac{1}{T})}{T} {T}^{\rm\alpha-1} dT+\int_{0 }^{+\infty } \frac{f(\frac{1}{T})}{T}T^{\rm\alpha-1} dT\\
=(-1)^{-\rm\alpha}\mathcal M\bigg\{ \frac{f(\frac{-1}{T})}{T}\bigg\}+ \mathcal M\bigg\{ \frac{f(\frac{1}{T})}{T}\bigg\}
\end{split}
\end{equation}
The normalized dimensional transformation ($\mathcal{NDT}$) of the second type is defined as:
\begin{equation}
\mathcal{NDT}\{f(t)\}:=F(\alpha):=\int_{-\infty}^{+\infty} f(t) \frac{t^{-\rm\alpha}}{\Gamma(1-\alpha)} dt 
\end{equation}
\\
where -$\alpha$=$\Re\{s\}-1$ ($0\leq \alpha\leq1$ ) and   s=$\Re\{s\} $ ($0\leq\Re\{s\}\leq 1 $) and 

Approach I:
\begin{equation}
\begin{split}
F(\alpha)=\frac{(-1)^{\rm\ s-1}}{\Gamma(1-\alpha)}\mathcal M\big\{ f(-t)\big\}+ \frac{1}{\Gamma(1-\alpha)}\mathcal M\big\{ f(t)\big\}
\end{split}
\end{equation}
Hint: $\alpha$ can be variable order and follows a trajectory.\\

\section{Mathematical Outlook and Future Work: Functional-Analytic Reconstruction}

The present work develops a geometric framework in which chord slopes are
realized by fractional operators of variable order. A natural next step is to
invert the perspective: instead of determining $\alpha(t)$ from $f$, one asks
whether an admissible order-function $\alpha(\cdot)$ can serve as a primary
descriptor from which $f$ can be reconstructed. This leads to a functional-analytic
program based on Volterra integral equations.

\subsection{Reconstruction map \texorpdfstring{$\mathcal R:\alpha\mapsto f$}{R: alpha->f}}
Fix an operator family $O^\alpha$ (e.g.\ Caputo for $0<\alpha<1$ or RL integrals for $\alpha>0$),
and consider the model relation
\begin{equation}\label{eq:outlook-model}
(O^{\alpha(t)} f)(t)=\Psi(\alpha(t)), \qquad t\in[t_0,T], \qquad f(t_0)=0,
\end{equation}
where $\Psi$ is a prescribed monotone function. For the Caputo case with $0<\alpha(t)<1$,
writing $y=f'$ yields an Abel-type Volterra equation of the first kind,
\[
\frac{1}{\Gamma(1-\alpha(t))}\int_{t_0}^{t}(t-s)^{-\alpha(t)}y(s)\,ds=\Psi(\alpha(t)).
\]
First-kind Volterra equations are typically ill-posed unless additional structure is imposed.
A standard approach is to \emph{regularize} the equation into a Volterra equation of the second kind.
One route (``Fix 1'') applies a suitable order-dependent operator (a variable-order analogue of a left inverse),
which in the constant-order setting reduces to the identity but in the variable-order case produces correction
terms involving $\alpha'(t)$ and logarithmic kernels. The result takes the schematic form
\[
y(t)=g_\alpha(t)+\int_{t_0}^{t}\widetilde K_\alpha(t,s)\,y(s)\,ds,
\]
where $\widetilde K_\alpha$ is integrable and proportional to $\alpha'(t)$.
This second-kind Volterra formulation is well-posed and admits existence, uniqueness, and stability
via Banach fixed-point arguments on short intervals, followed by continuation.

\subsection{Inverse order map \texorpdfstring{$\mathcal T:f\mapsto \alpha$}{T: f->alpha}}
A complementary direction is to define an order map implicitly by solving, for each $t$,
\[
\Phi_{f,t}(\alpha):=(O^\alpha f)(t)-\Psi(\alpha)=0.
\]
Under monotonicity and range conditions (analogous to those used in the geometric part of this paper),
one expects unique solvability of $\alpha_f(t)$ and regular dependence on $t$.
Establishing compatibility between $\mathcal R$ and $\mathcal T$ (e.g.\ $\mathcal T(\mathcal R(\alpha))=\alpha$
on an admissible class) would provide a rigorous order--function representation theory.

\subsection{Relation to distributed-order models}
Distributed-order fractional operators have the form
\[
\int_A \omega(\beta)\,O^\beta f\,d\beta,
\]
which represent a spectrum of memory scales. The variable-order setting corresponds formally to a
time-localized weight $\omega(\beta,t)=\delta(\beta-\alpha(t))$, suggesting approximation and model-reduction
links between distributed-order and variable-order dynamics. A systematic study of these connections,
including error bounds and identifiability of $\alpha(t)$, is a promising topic for future work.

\end{document}